\documentclass[conference]{sig-alternate-10pt}

\usepackage{algorithm2e}
\usepackage[tight,footnotesize]{subfigure}
\usepackage{multirow}
\usepackage{booktabs}
\usepackage{float}
\usepackage[usenames, dvipsnames]{color}
\usepackage{comment}
\usepackage{rotating}
\usepackage{amssymb}
\usepackage{todonotes}
\usepackage{hhline}
\usepackage{lastpage}

\usepackage{caption}
\DeclareCaptionType{copyrightbox}

\DeclareMathOperator*{\argmin}{arg\,min}

\newcommand{\red}[1]{\textcolor{red}{#1}}
\newcommand{\mycomment}[2]{\textcolor{red}{#1:~#2}}

\newcommand{\lw}[1]{\mycomment{\textcolor{blue}{LW}}{\textcolor{blue}{#1}}}

\newcommand{\smell}[0]{...\,}

\usepackage[
  pass,% keep layout unchanged 
]{geometry}

\usepackage{enumitem}
\setlist[itemize]{noitemsep, topsep=4pt, leftmargin=*}
\setlist[enumerate]{noitemsep, topsep=4pt, leftmargin=*}

\begin{document}
\title{XBF: Scaling up Bloom-filter-based Source Routing}
%%%\title{Bloomsday: Scaling up Bloom-filter-based Source Routing using Traffic-Aware Graph Partitioning}

%\numberofauthors{4}
%\author{
%\begin{tabular*}{0.7\textwidth}{cc}
%Markku Antikainen & Liang Wang \\
%\affaddr{Aalto University} & \affaddr{University of Cambridge} \\
%\emm{markku.antikainen@aalto.fi} & \emm{liang.wang@cl.cam.ac.uk} \\
%& \\
%Dirk Trossen & Arjuna Sathiaseelan \\
%\affaddr{InterDigital Europe, Ltd} & \affaddr{University of Cambridge} \\
%\emm{dirk.trossen@interdigital.com} & \emm{arujuna.sathiaseelan@cl.cam.ac.uk}
%\end{tabular*}
%}

\author{
\alignauthor Markku Antikainen\textsuperscript{*}, Liang Wang\textsuperscript{\dag}, Dirk Trossen\textsuperscript{\ddag}, Arjuna Sathiaseelan\textsuperscript{\dag} \\
\affaddr{\textsuperscript{*}Aalto University, \textsuperscript{\dag}University of Cambridge, \textsuperscript{\ddag}InterDigital Europe, Ltd} \\
\email{first.last@\{aalto.fi, cl.cam.ac.uk, interdigital.com\}}
}

\maketitle

\begin{abstract}
A well known drawback of IP-multicast is that it requires per-group state to be stored in the routers. 
Bloom-filter based source-routed multicast remedies this problem by moving the state from the routers to the packets. However, a fixed sized Bloom-filter can only store a limited number of items before the false positive ratio grows too high implying scalability issues. Several proposals have tried to address these scalability issues in Bloom-filter forwarding. These proposals, however, unnecessarily increase the forwarding complexity. 

In this paper, we present Extensible-Bloom-filter (XBF), a new framing and forwarding solution which effectively circumvents the aforementioned drawbacks. XBF partitions a network into sub-networks that reflect the network topology and traffic patterns, and uses a separate fixed-length Bloom-filter in each of these. We formulate this partition assignment problem into a balanced edge partitioning problem, and evaluate it with simulations on realistic topologies. Our results show that XBF scales to very large networks with minimal overhead and completely eliminate the false-positives that have plagued the traditional Bloom-filter-based forwarding protocols. It furthermore integrates with SDN environments, making it highly suitable for deployments in off-the-shelf SDN-based networks.

\end{abstract}

\section{Introduction}
\label{sec:intro}

Conventional IP routing has been suffering from state explosion at the network core. It is well known that the backbone routers have to maintain routing tables with tens of thousands of entries.
This not only introduces a significant amount of management overhead, but also hinders the scalability of IP-routing and IP-multicast in particular. The issue is exacerbated by the emergence of
Internet-of-Things (IoT) paradigm, where an even larger number of devices are expected to be connected over IP-enabled networks, creating even more state in the network. Furthermore, the routing
operations in IP networks require advanced memory and packet switching technology in order to implement the longest-prefix matching necessary to forward packets from the sender to the receiver.
 
One approach to address these pinching points of IP routing is that of Bloom-filter-based source-routing\footnote{While work on BF forwarding does not prevent its application for inter-domain routing, most solutions have been applied to intra-domain routing only. We shall therefore limit our presentation to the intra-domain problem, too.}. In these schemes, the path from sender to receiver is computed by a dedicated
path computation element, while each link along the path is identified with a domain-unique link identifier~\cite{Jokela:2009, 6503540}. Bloom filters are used to encode all link identifiers along the path from the sender to the receiver(s) within a single bitfield~\cite{Jokela:2009}. With that, multicast is naturally supported while the forwarding operation becomes that of a membership test in a bitfield (the Bloom filter). 
%In a further simplification of the BF forwarding solution in \cite{Jokela:2009}, each link identifier can be represented as a uniquely assigned bit within a bitfield of well-defined length. With that, any bit entry 1 in the bitfield indicates the necessary forwarding along the link represented by said bit. 
In other words, the forwarding decision at each switch can be realized as a simple bitwise AND/CMP operation, compared to longest prefix matching in IP routers. Applications for such simple forwarding solution can be found in
multi-site VPN solutions~\cite{zahemszky2010mpss,7152107}, in datacenters~\cite{5700387,6045301}, and in information-centric networking (ICN) architectures~\cite{fotiou2012developing}. Within these applications, the need for centralized path computation is well aligned with the overall network architecture, while a minimal forwarding operation is specifically
aimed at for performance reasons. Furthermore, work in \cite{point} has shown that BF-based forwarding can be easily implemented in off-the-shelf SDN equipment at the cost of constant-sized flow matching tables, making it highly suitable for deployments in such emerging network environments.
 
%However, the simple bitfield approach of BF forwarding comes with a significant scalability drawback. The size of the bitfield that is used for forwarding (and which holds the path information) determines the maximum number of links being supported and therefore maximum size of the overall network topology. Approaches presented in~\cite{Jokela:2009} and applications such as those presented in~\cite{zahemszky2010mpss,5700387,fotiou2012developing} usually rely on bitfields of

However, BF forwarding comes with a significant scalability drawback. The size of the bitfield that is used for forwarding (and which holds the path information) determines the maximum number of links that can be stored on it. Thus, the size of the bitfield defines the maximum size of the network topology and the maximum size of the multicast group.
Approaches presented in~\cite{Jokela:2009} and applications such as those presented in~\cite{zahemszky2010mpss,5700387,fotiou2012developing} usually rely on bitfields of 256 bit length, which limits their deployment capability. Extending the bitfield size is surely possible but only within the limit of defining a larger bitfield, requiring the change of all protocol headers, which is
hardly a suitable approach for flexible deployment across a range of topologies.
 
In order to better frame the particular design choices we make for our solution to this scalability problem, we formulate the following main goals for a suitable BF forwarding solution:
\begin{enumerate}
\item Any solution must support any topology and group size: in contrast to existing solutions \cite{Jokela:2009,6503540}, we are seeking a solution that does not limit the topology sizes supported.\footnote{In environments where we tolerate false positives in BF-based forwarding, this requirement addresses the need for supporting arbitrary topology sizes with defined false positive tolerance boundaries.}
\item Any solution should allow for instantaneous multicast group formation: one aspect of BF forwarding is that multicast trees can easily be formed by unicast path information by simply bitwise ORing the
unicast BF information into a single (now multicast) BF identifier. Any solution should preserve this capability in order to allow for applications such as~\cite{trossen2015ip}.
\item The basic forwarding operation should be kept as simple as possible to the original AND/CMP based forwarding, i.e., an extensible BF scheme should not increase individual forwarding costs.
\item Any solution should be easily realizable over an SDN-based infrastructure in order to ensure easy deployment with such emerging network environments. Solutions such as~\cite{point} have outlined how to implement basic BF forwarding over SDN equipment. Any solution should preserve such capability as much as possible.
\end{enumerate}
 
In this paper, we propose the Extensible Bloom-filter Forwarding (XBF) solution with the attempt to address the requirements outlined above. For this, we take a partitioning-based approach in which we partition the network into
several parts (i.e., sub-network) so that each part contains an equal amount of links. The forwarding inside every partition happens with a standard fixed-length Bloom-filter forwarding. When a packet traverses from one
partition to another, the corresponding partition-specific Bloom-filter must be loaded into the portion of the header that is used for BF matching; this copy operation causes some processing overhead. In order to reduce
this processing overhead, XBF incorporates an intelligent partition algorithm which is able to exploit both network topological properties and traffic patterns, so that inter-partition traffic volume is
minimised. Moreover, the resulting traffic-aware partitioning can also improve header overhead by effectively reducing the number of Bloom-filters carried in a packet. For this, we formulate the XBF design problem into a balanced edge partition problem, and present an intelligent partition algorithm that adopts the traffic-aware weighting scheme in order to minimise both header and processing overhead.
 
The rest of the paper is organised as follows. Section 2 presents a more detailed background on BF-based forwarding. We then describe the aforementioned zoning approach to BF forwarding as a system model in Section 3, followed by our traffic-aware partitioning approach in Section 4. In order to verify the compliance with our requirements 3 and 4, we present
the experimental setup for our evaluation in Section 5. Our results regarding header and processing overhead are presented in Section 6, together with results on impact of different traffic patterns. As a proof-of-concept, specifically addressing requirements 3 and 4 again, we also present an implementation of our approach in Section 7, using a P4 software switch~\cite{bosshart2014p4}. The paper then concludes with a discussion on possible future work.

\section{Background}
\label{sec:background}

\subsection{Bloom-filter forwarding}

Bloom-filter forwarding is an attractive scheme especially for content distribution applications due to two reasons: (1)~unicast and multicast both use the same forwarding operation; (2)~multicast does not require per-flow or per-group state to be stored in the network. These advantages are in stark contrast to the conventional IP-based multicast solutions where the routing table sizes grow as a function of both network size and the number of  multicast groups in the network. 

The basic idea behind Bloom-filter forwarding is simple. Every unidirectional link in the network is given a unique \emph{link identifier} $l$, which is a $m$-bit long string with $k$ independently selected random bits set to one (i.e., overlapping is possible). By so doing, any delivery path $L$ can be described as a set of links that a packet will traverse in the network. 

\begin{figure}[t]
%\missingfigure[figwidth=1\columnwidth]{Basic BF-operations and false positive example.}
\includegraphics[trim = 4mm 4mm 4mm 4mm, clip, width=1\columnwidth]{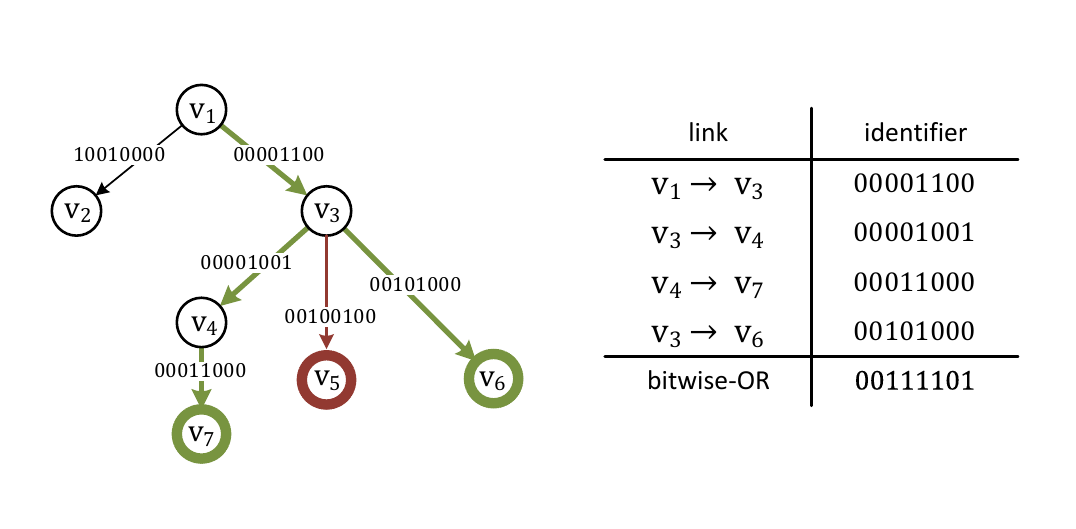}
\caption{Bloom-filter for sending traffic from $v_1$ to receivers $v_6$ and $v_7$ causes a false positive and makes the packet to traverse also the link $v_3 \rightarrow v_5$.}
\label{fig:bfBasics}
\end{figure}

In general, Bloom filter routing can be split into two parts: \emph{path construction} and \emph{packet forwarding}. In the path construction phase, the sender or a logically centralized topology manager creates an in-packet Bloom-filter by combining the link identifiers towards every recipient together. In Figure~\ref{fig:bfBasics}, the multicast tree to $v_6$ and $v_7$ contains four links (marked as green). The in-packet Bloom-filter $F$ is created by simply bitwise-ORing the link identifiers together, namely $F := \bigvee_{\forall l \in L}^{}{l}$, where $F$ is the in-packet Bloom-filter, $L$ is the set of link identifiers on a path, and $\vee$ is the bitwise-OR operation Then, $v_1$ embeds the resulting filter \texttt{00111101} into the packet header and sends it out to the network.

%In general, Bloom filter routing can be split into two parts: \emph{path construction} and \emph{packet forwarding}. In the following, we explain these using the situation presented in Figure~\ref{fig:bfBasics} as an example.

%Every unidirectional link in the network is allocated with a unique identifier $l$, which is a $m$-bit long string with $k$ independently selected random bits set to one (i.e.\ overlapping is possible). By so doing, any delivery path can be described as a set of links $P$ that a packet will traverse by after entering into the network. For example, four links are involved in the multicast tree to deliver a packet to $v_6$ and $v_7$, as listed in the left column of the table, also marked with red color accordingly in Figure~\ref{fig:bfBasics}. The content provider, i.e., $v_1$ in Figure~\ref{fig:bfBasics}, is aware of all the subscribers hence is responsible for path construction. The content provider embeds all the links into a Bloom filter $F$ by simply bitwise-OR every link identifier together, namely $F := \bigvee_{\forall l \in P}^{}{l}$, where $F$ is the in-packet Bloom-filter, $P$ is the set of link identifiers on a path, and $\vee$ is the bitwise-OR operation, as described above. Then $v_1$ embeds the result \texttt{00111101} into the packet header and sends it out to the network.

The packet forwarding step becomes rather straightforward with a packet header containing all the links which need to be traversed by. Whenever a packet arrives at a switch, the switch extracts the Bloom-filter from the packet header, and for every outgoing link, it checks whether the particular link identifier has been embedded in the Bloom-filter. This operation, referred to as \textit{membership test}, is done by checking the equality $F \wedge l \stackrel{?}{=} l$, where $l$ is the link identifier, and $\wedge$ denotes the bitwise-AND operation.

While Bloom-filter membership test makes the forwarding very simple, it also makes forwarding operation susceptible to false-positives. Given a fixed-length Bloom-filter, as more and more links are embedded in, the membership test may mistakenly report the presence of a link whose identifier actually has never been embedded before. Such error is specifically referred to as \textit{false positive}. False positive indicates that a switch may forward a packet to an interface where the packet is not intended to. For example in Figure~\ref{fig:bfBasics}, a packet with Bloom-filter \texttt{00111101} will be forwarded also to $v_5$ even though it is only intended to $v_6$ and $v_7$. The reason is because the membership test for link $v_3 \rightarrow v_5$ produces false positive, i.e., \texttt{00111101}~$\wedge$~\texttt{01001000}~$=$~\texttt{01001000}.

False positive rate (FPR) is a widely used metric to evaluate a Bloom-filter-based routing scheme, which effectively tells the probability that a switch forwards packet to an unintended interface.
FPR is a function of the Bloom-filter parameters (i.e.,\ values of $m$ and $k$) and the number of link identifiers stored in the filter. It is obvious that FPR will be high for large multicast trees, since more links are involved in delivery and their identifiers have to be embedded. High FPR values can cause a significant amount of redundant traffic increasing the probability of congestion in the network. Another problem with false positives is that they may cause a packet to be forwarded back to previously visited nodes in the multicast tree and thus render (possibly infinite) forwarding loops.

We now continue by describing how earlier research has approached these challenges.

%The most effective way to keep FPR low is either increasing the length of Bloom-filter or decreasing the number of link identifiers to be embedded. Both, unfortunately, are problematic from the scalability point of view. For example, dynamically varying the length of a Bloom-filter (according to the number of links in a multicast tree) increases the forwarding complexity compared to the fixed-length Bloom-filters where forwarding can be done with simple bitwise operations. This is because using variable-length Bloom-filters requires the switches to dynamically calculate the link identifiers for Bloom-filters of different lengths. To avoid dynamic calculations, although switches can store pre-calculated link identifiers for different lengths, it inevitably increases the amount of states in the switches by orders of magnitude.

%%% Our proposal avoids these problems by intelligently partitioning the network into equisized parts inside which the forwarding uses a fixed-sized Bloom-filters, as described below.

\subsection{Challenges of BF-forwarding}
\label{sec:relatedWork}
%\as{I would actually merge the related work with the background section!}

%\red{Making Bloom-filter forwarding scale -- other approaches}
% MSBF: Requires that the link-identifiers can be calculated line-speed (minor drawback only, though) 
% MSBF: Bloom-filters of different lengths cannot be combined!

The scalability-issues caused by the fixed Bloom-filter length are well known and have been identified already in early proposals such as Free-Riding Multicast~\cite{Ratnasamy:2006} and LIPSIN~\cite{Jokela:2009}. In general, there are three approaches that can be found from the literature regarding increasing the scalability of Bloom-filter source routing. The first, and simplest way to achieve scalability is to disallow too large multicast-trees and instead use several trees when serving content to a very large audience~\cite{6364409,Nikolaevskiy201579}. While the solution is very simple and effective, it naturally increases the amount of traffic in the network and is suitable only in fairly small networks.

The second way to improve the scalability of Bloom-filter forwarding is to use some method to decrease the FPR. The idea is that when the FPR gets smaller, a fixed sized Bloom-filter can store larger multicast trees. %There are several methods how the FPR can be reduced. 
One commonly used way to reduce FPR is to use several coexisting link-identifiers for every link: when the multicast trees are constructed, the link identifiers are chosen so that the number of false positives is minimized~\cite{Jokela:2009,Hao:2007}. The drawback of this approach is it increases the amount of state stored in the switches. Also, these approaches only slightly decrease the FPR and do not provide scalability for arbitrary large networks. 
Furthermore, because the FPR still remains positive, these forwarding schemes must employ extra countermeasures against forwarding loops caused by false-positives thus further complicating the router design~\cite{5935060,6892989}. 

The third way to solve the scalability problem is to encode the multicast-trees into variable sized Bloom-filters~\cite{5935060,6877748}. This basically means, that when a multicast tree grows too large (and thus causes too may false positives) the in-packet Bloom-filter is dynamically extended to keep the FPR acceptable. This requires that there is an additional length-field in the packet header which tells the length of the packet's Bloom-filter. These variable-length Bloom-filter solutions would appear palatable due to the fact that it scales to indefinitely large networks and multicast groups. 

Making the Bloom-filters variable-length has, however, two major drawbacks. First, it makes the multicast-group management more difficult. This is because vari\-able-length Bloom-filters cannot be combined together simply by bitwise ORing them together (they are of different lengths, after all). Because of this, the topology manager becomes a bottleneck as it must be consulted whenever the multicast tree changes (i.e., a node joins or leaves the multicast group). The second problem is that variable-length Bloom-filters vastly increase the complexity of the forwarding. When the Bloom-filter lengths vary from packet to packet, the link identifiers have to be calculated based on the length of the Bloom filter. Thus, the switches have to calculate link-identifiers at line-speed. Alternatively, the switches have to store pre-calculated link-identifiers for every possible Bloom-filter length. This, on the other hand, would significantly increase the amount of state stored in the network.

There are also proposals that try to combine the aforementioned approaches. For example, Multi-stage Bloom-filter (MSBF) tries to reduce the false positives by splitting the multicast-tree into several variable-length stages which use different Bloom-filters~\cite{6503540, 6655118, 6877748}. The problem with this kind of solutions is, again, that they tend to make group management very difficult and increase the complexity of the forwarding nodes.

\section{XBF forwarding scheme}
\label{sec:forwarding}

In this section, we introduce the extensible Bloom-filter forwarding (XBF) and the system model under which it operates. XBF addresses the fundamental scalability issue of conventional Bloom-filter forwarding by partitioning a network into equisized parts inside which a separate Bloom filter is constructed.

\begin{figure*}[t]
\centering
\subfigure[A sample topology with three partitions, popping switches are $v_1$, $v_2$, and $v_3$ at the boundary of partitions.]{
	\includegraphics[trim = 4mm 4mm 118mm 4mm, clip, width=0.45\textwidth]{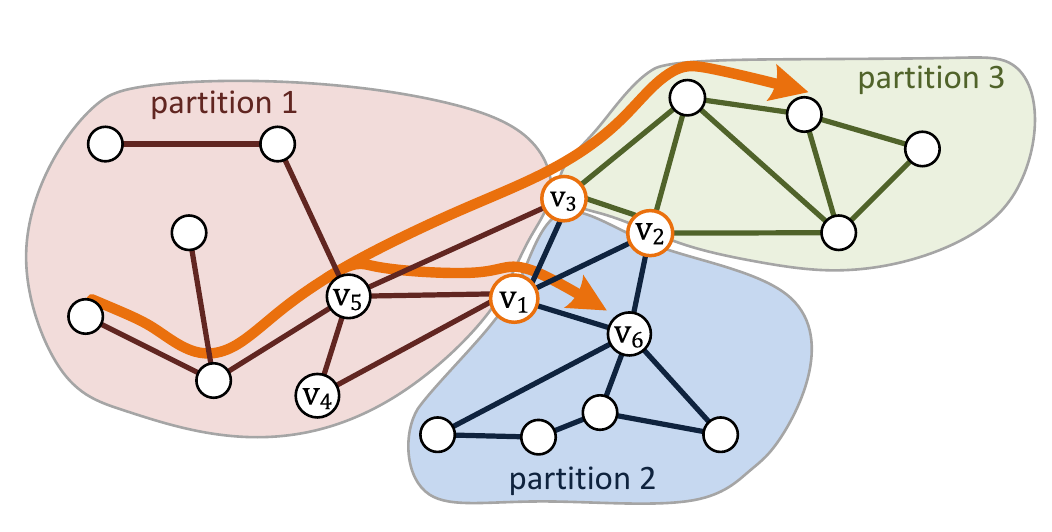}
}\qquad
\subfigure[The packet header format and the popping operation at $v_1$ when the packet traverses the path ${v_5 \rightarrow v_1 \rightarrow v_6}$.]{
	\includegraphics[trim = 110mm 4mm 4mm 4mm, clip, width=0.45\textwidth]{figures/system_model.pdf}}\vspace{-0.5\baselineskip}
\caption{XBF system model, a naive multicast tree is used to illustrate header format and popping operation.}\label{fig:system}
\end{figure*}

\subsection{System Model}
\label{sec:system}

XBF forwarding scheme is designed for wide scale intra-domain multicast. The scheme works in networks where roughly half of the switches can perform packet matching on arbitrary fields (e.g.\ software switches or P4-switches) while  the rest of the switches can be regular OpenFlow-switches.
%  where majority of the switches are programmable (e.g.\ software switches or SDN switches). 

We assume a network that can be represented as a directed graph, $G_n = (V_n, E_n)$, where $V_n$ is the set of switches and $E_n$ is the set of links\footnote{In this paper, we use the terms \emph{node} and \emph{switch} interchangeably as well as the terms \emph{edge} and \emph{link}.}. The basic idea in XBF is to group the network edges into partitions so that every partition contains at most 256 edges. Then, every partition is allocated with a separate 256-bit partition Bloom-filter. Because the partition Bloom-filters are 256-bit long and every partition contains at most 256 edges, we can assign a \emph{unique one-bit link identifier} for every link instead of creating the link identifiers randomly. Therefore we can completely get rid of false-positives.\footnote{Technically, the described data-structure is no longer a Bloom-filter. We nevertheless abuse this term as the operations done with these one-bit identifiers match the Bloom-filter operations.} 
The maximum partition size is chosen as 256 because it fits into the source and destination address fields of a IPv6 header.

Figure~\ref{fig:system} presents the XBF packet format. The packet header has two parts: \textit{iBF} contains the Bloom-filter used for intra-partition forwarding, and \textit{zBF} contains the Bloom-filters of all of the partitions which the packet needs to traverse by. The content in zBF remains the same throughout a delivery whereas the content in iBF changes dynamically depending on which partition the packet is currently in. Together the iBF and zBF constitute the header overhead.

There are three main entities in a XBF-enabled network: \textit{topology manager}, \textit{popper switches}, and \textit{forwarder switches}. 
%For the ease of description, we choose to follow a LIPSIN-like~\cite{Jokela:2009} model. Nonetheless, it is worth noting that the proposed scheme is by no means constrained by ICN but can be applied to any general Bloom-filter based routing schemes. Figure.~\ref{fig:system} depicts the overall system.
The topology manager is a logically centralised entity similar to a SDN controller, with three main functions:
(1)~it partitions the network so that each partition contains at most 256 directed links; (2)~it assigns the one-bit link identifiers for every link in the network; (3)~it creates Bloom-filters that can be used to send packets from a source to a destination. 
Note that (1) and (2) are only performed once for the whole system while (3) is performed once for a new multicast tree.

A popper switch (or popper) refers to a switch sitting at the boundary of partitions (e.g., $v_1$, $v_2$, and $v_3$ in Figure~\ref{fig:system}). More specifically, a switch $v$ is a popper iff the edges incident to $v$ belong to at least two partitions. Meanwhile, a forwarder switch (or forwarder) refers to a regular switch with all its links allocated to the same partition. 
Forwarder switch is a simple device that only performs basic Bloom-filter forwarding operations -- it checks which of its next-hop links are included in the iBF then forwards the packets to the corresponding interface. 
This simple operation, however, only applies to the forwarding occurs inside one partition. If a packet is intended to another partition, it must go through a popper which lies at the border of multiple partitions.
Unlike the basic forwarding, a popper copies a partition Bloom-filter from the zBF to the iBF every time it performs the membership tests for the links of the partition. We refer this operation as \emph{popping} in the paper.

Figure~\ref{fig:system} illustrates popping operations. The multicast tree (marked as orange arrows) crosses partition borders at two different popping nodes: $v_1$ and $v_3$. When the packet arrives at $v_1$, the popper has to pop (i.e., copy) the blue Bloom-filter from zBF to iBF. After popping, $v_1$ can check to which outgoing blue interfaces the packet should be forwarded as a regular forwarding element. However, the operation is slightly different at popper $v_3$ who has to pop the packet twice. The reason is that $v_3$ sits at the border of three partitions hence it needs to perform the popping for both green Bloom-filter (partition 3) and blue Bloom-filter (partition 2). As we can see, the number of popping operations a popper needs to perform for a packet depends on the number of incident partitions.

%%% We now move to consider how the way the network is partitioned can affect the number of popping operations. 

\begin{comment}
\begin{figure}[t]
\centering
\includegraphics[trim = 4mm 4mm 4mm 4mm, clip, width=1\columnwidth]{figures/popper_and_packet.pdf}
\caption{Popper switch $v_p$. The colours denote the partition to which each link belongs. \lw{Later, we can add some explanatory text in the figure to utilise the space mroe efficiently.}}\label{fig:popper}
\end{figure}
\end{comment}

\addtolength{\tabcolsep}{-0.2mm}
\begin{table}\centering
\scriptsize
\begin{tabular}{ p{1.9cm} | p{6cm} }
\hline
$F$ & in-packet Bloom-filter, $F \in \{0,1\}^m$ \\
$m$ & length of a Bloom filter \\
$\wedge$, $\vee$ & bitwise-AND and bitwise-OR respectively \\
$G = (V, E)$ & directed graph with vertices $V$ and edges $E$ \\
$G_n = (V_n, E_n)$ & directed network graph \\
$G_c = (V_c, E_c)$ & directed connectivity graph, $V_c = E_n$ \\
%$\tau_{a}$ & traffic volume on node $v_a$ \\
$\tau_{a,b}$ & traffic volume on network link $v_a \rightarrow v_b$ \\
$\nu$ & imbalance coefficient in vertex partitioning \\
$P$ & set of partitions \\
$p_e$ & partition to which a link (or connectivity graph vertex) $e$ belongs to, $p_e \in P$ \\
$N(v)$ & partitions that are adjacent to node $v$, $N(v)~\subseteq~P$ \\
$\phi_e$ & number of popping operations done on each packet traversing link $e$ \\
\hline
\end{tabular}
\caption{Notation used in this paper}
\label{tlb:notation}
\end{table}

\subsection{Optimising Popping Via Edge Partition}
\label{sec:optimise}

From both network operator's and end user's perspective, the total number of poppings in the network needs to be minimised, which requires the topology manager to strategically select poppers to connect different partitions.
To help us in formulating the optimisation problem in Section.~\ref{sec:partitioning}, we first define the required number of popping operations as follows. Let $p_e$ denote the partition which a directed link $e$ belongs to. Let $H_e$ be the set of possible next-hop links that the packets can traverse after link $e$. The following equation gives the number of poppings required for each packet traversing via $e$ (note that the set cardinality tells the number of \emph{distinct} elements in it).
\begin{equation}
\phi_e = \Big| \Big(\bigcup_{e' \in H_e} \{p_{e'}\} \Big) \setminus \{p_e\} \Big|
\label{eq:noPoppings}
\end{equation}

%The union in Equation~\ref{eq:noPoppings} creates a set of all possible next-hop partitions to which a packet can traverse after edge $e$. 

There are two technical details worth mentioning: First, the set of possible next-hop links $\phi_e $ excludes the interface where a packet originally came from in order to avoid loops. Second, the popping operation is not necessarily needed for all passing-by packets. A popper only performs popping for the packets entering into different partitions, i.e., cross-partition traffic. For the packets remaining in the same partition after forwarding, popping operation can be skipped. We will present the actual implementation details in Section~\ref{sec:sdn}.

%%%Second, the popping needs to be performed for all passing-by packets even though some may not actually leave the partition. The reason is that a popper does not know whether a packet is intended to another partition without a membership check. 

%%%\lw{[Is this still valid given your P4 implementation???]}
%%%We now describe an algorithm that optimises the partitions based on the network topology and traffic volumes.

%It has been shown that edge partition and vertex partition are equivalent~\cite{bourse2014balanced}. Therefore we first translate a normal graph into connectivity graph. The obvious benefit is that we can directly apply classic vertex partition algorithm without any modification. Another implicit benefit is due to the feature of connectivity graph. \lw{Present the example here.} We can reduce the probability of loops to some extent. We are able to handle asymmetric traffic so TM can construct a better distribution tree (i.e., with lower popping operation on average???). The feature guarantees that connectivity graph is at least as good as naive translation.

\section{Traffic-Aware Partitioning}
\label{sec:partitioning}

We have designed Jigsaw, an algorithm to calculate the optimal partitioning of a given network to minimise poppings. Jigsaw is traffic-aware and is also an important component in topology manager, playing a significant role in XBF performance.

In brief, Jigsaw aims to minimise the traffic volume across partition boundaries (i.e., inter-partition traffic) in order to minimise the total number of poppings. 
More specifically, for a given traffic pattern and a network topology, Jigsaw optimises the partitions boundaries so that majority of the traffic occurs within a single partition, while simultaneously keeping the number of links in each partition constant (or near-constant).
There are three steps in the algorithm:
\begin{enumerate}
\item \textbf{Weighting}: the algorithm adds weight to every link in the network based on the actual traffic volumes.
\item \textbf{Translation}: the algorithm transforms the original network graph into a \emph{connectivity graph}.
\item \textbf{Partitioning}: the algorithm performs balanced vertex partitioning on the connectivity graph to obtain an optimal partitioning.
\end{enumerate}
We now separately describe the details of these steps.

\subsection{Embedding weights in original network}
\label{sec:weighting}

The very first step in Jigsaw is assigning weight on every directed link according to the traffic volume passing by. Since forwarding and popping operations are performed on per-packet basis, we set the weight of a link equal to the number of packets passing through in a well-defined time-unit (e.g., an hour).

Although weighting indicates that a topology manager needs to possess the global knowledge of traffic volume in a network, if such information is not available, the traffic volumes can be approximated with switches' betweenness centrality values. Many prior work have shown that there is a strong correlation between traffic patterns and networks' topological properties~\cite{Mahadevan:2006:IAT:1111322.1111328, Borgatti200555} 
%%%\ma{Liang, do you know a citation to put here?}.

%Although we assume that the topology manager has the global knowledge of the network, the traffic information needed in weighting can also be measured in a straightforward way. Even when such measurement is expensive or impossible, Jigsaw is still able to estimate the traffic pattern by using switch's betweenness centrality values. Many prior work have shown that there is a strong correlation between traffic pattern and network topological properties~\cite{???}.

\subsection{Constructing a connectivity graph}
\label{sec:connectivity}

\begin{figure}[t]
\centering
\includegraphics[trim = 5mm 5mm 5mm 5mm, clip, width=1\columnwidth]{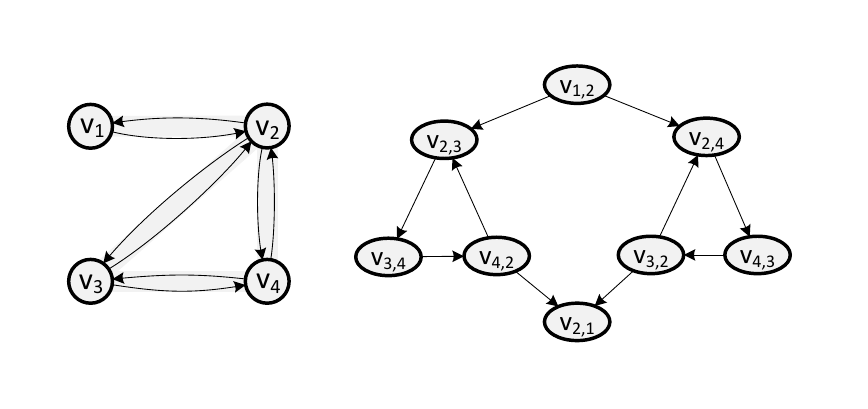}
\caption{A network graph and a connectivity graph created from it. Link weights (network graph) and vertex weights (connectivity graph) are omitted.}\label{fig:connectivityGraph}
\end{figure}

The second step is to construct a connectivity graph~\cite{6616021} from the original network. The algorithm for creating such a connectivity graph is simple. Firstly, for each directed \emph{link} $(v_i, v_j)$ in the original graph, a matching \emph{vertex} $(v_i, v_j)$ is added to the connectivity graph. Accordingly, the original link weight becomes the vertex weight in the connectivity graph. Then edges are added to the connectivity graph with a simple rule: a directed edge is added between two connectivity graph vertices iff a packet can traverse from the corresponding network link to another in the original graph. Figure~\ref{fig:connectivityGraph} illustrates a connectivity graph created from a simple network graph. 

There are several motivations of constructing a connectivity graph. First, we can translate a balanced edge partition problem to an equivalent balanced vertex partition problem. While the two problems have been shown to be equivalent~\cite{bourse2014balanced}, balanced vertex partitioning is already well studied hence there are many mature algorithms\footnote{To be more specific, the edge-partition problem with \emph{no aggregation} is equivalent to a balanced vertex-partition problem. This corresponds to the problem we are solving.}.

Second, a connectivity graph is able to reflect actual paths more accurately and further remove many potential loops. For example in Figure~\ref{fig:connectivityGraph} where a packet traverses from $v_3$ to $v_2$, it is obvious that the next-hop link cannot be $v_2 \rightarrow v_3$ since that is where the packet comes from. The directed network graph does not encompass this information about physical interfaces, whereas connectivity graph does not have this limitation -- an edge between two connectivity graph vertices means by definition that a packet can follow that path.

Third, a connectivity graph is able to take traffic direction into account, which can lead to better partitioning especially when traffic is asymmetric in different directions. This can be seen by examining the connectivity graph in Figure~\ref{fig:connectivityGraph}: the connectivity graph seems to contain two parts that are very weakly connected to each other (i.e. $v_{2,3} \rightarrow v_{3,4} \rightarrow v_{4,2}$ and $v_{2,4} \rightarrow v_{4,3} \rightarrow v_{3,2}$). This hints that an optimal partitioning may, rather counter-intuitively, may place reverse links to different partitions. This is also what we observed when partitioning realistic network graphs (see Section~\ref{sec:evalResults}). 
An edge partitoining done directly on the network graph could not 
%We show later in Section~\ref{sec:evaluation} that this, indeed, is the case.

%%% \lw{Fourth, can connectivity graph help in partitioning heterogeneous network?}

\begin{algorithm}[t]
\SetAlgoLined
\KwData{Network graph $G_n = (V_n, E_n)$. Every $e \in E_n$ is a 3-tuple $(v_a, v_b, \tau_{a,b})$, where $v_a$ and $v_b$ denote the link start and endpoints respectively, and $\tau_{a,b}$ denotes the traffic volume on that link.}
\KwResult{Partitioned graph}
\small{
	\tcp{Create connectivity graph and annotate it with link weights}
	\ForEach{ $(v_a, v_b, \tau_{a,b}) \in E_n$ }{
		$V_c := V_c \cup \big\{(v_a, v_b, \tau_{a,b})\big\}$\;
		\ForEach{ $v_c \in$ getNeightbours($G_n, v_b$) }{
			\If{$v_a \neq v_c$}{
				$E_c := E_c \cup \Big\{\big((v_a, v_b, \tau_{a,b}), (v_b, v_c, \tau_{b,c})\big)\Big\}$\;
			}
		}
	}
	
	\tcp{Apply the vertex-partitioning algorithm}
	$G_c := (V_c, E_c)$\;
	 \textit{no\_partitions} $=\big( |E| / 256 \big) \times \nu$\; 
	\Return \textit{balancedVertexPartitioning($G_c$, \textit{no\_partitions})}
}
\caption{Jigsaw, the traffic-aware network partitioning algorithm}
\label{alg:TheAlgorithm}
\end{algorithm}

\subsection{Applying vertex partitioning}
\label{sec:partition}

By definition, the balanced edge partitioning problem in the original graph can be naturally solved by applying a balanced \emph{vertex} partitioning on the corresponding connectivity graph. The goal is to group the vertices of a given connectivity graph $G_c=(V_c, E_c)$ into $n$ components of at most 256 vertices, while trying to minimise a well-defined \emph{total cost} function.
%at most of size $\nu \cdot (|V|/k)$, meanwhile tries to minimise a well-defined \emph{total cost} function.

%Given every vertex $v \in V_c$ is annotated with traffic volume $\tau_v$, we let $p_v \in P$ denote the partitions which vertex $v$ belongs to. In addition, we let $N(v) \subseteq P$ denote the set of partitions which $v$'s neighbours belong to. The total cost is then defined as follows:

We use $\tau_v$ to denote the traffic volume on a connectivity graph vertex $v \in V_c$ (remember that every connectivity graph vertex uniquely describes a link in the physical network). 
Also, we use $p_v \in P$ to denote the partitions of $v$, and $N(v) \subseteq P$ to denote the set of partitions to which $v$'s neighbours belong to. 
The total cost of a partitioning is now defined as follows:
\begin{align}
\mathit{totalv} &= \sum_{v \in V} \tau_v \cdot | N(v) \setminus \{p_v\} | 
%\\ &= \sum_{v \in V} \tau_v \phi_v 
\label{eq:totalv}\end{align}
Comparing to Eq.~\ref{eq:noPoppings}, besides the fact that traffic volume has been taken into account, the only difference in Eq.~\ref{eq:totalv} is that the calculations are performed over all vertices in the connectivity graph (corresponding to the links in the original graph). Therefore, it is not difficult to realise that $\mathit{totalv}$ equals to the total number of popping operations in the network.
Consequently, the optimal partition can be obtained by finding a solution that minimizes the $\mathit{totalv}$:
\begin{equation}
\argmin_{\forall v \in V, p_v} \mathit{totalv}
\end{equation}

This optimization problem is well known and is called as graph partitioning while minimizing the total communication volume~\cite{hendrickson2000graph}. The problem has been shown as NP-hard. However, there are efficient approximation algorithms that can solve the problem in polynomial time given the condition for equisized partitions is slightly relaxed~\cite{graphPart2006}. 
In our case, since we only require that the partition sizes \emph{do not exceed} 256, we can simply apply, for example, a multilevel $k$-way partitioning algorithm to find a nearly-optimal solution for the problem~\cite{metis}. This algorithm is part of widely used vertex partitioning library called {\sc Metis}~\cite{karypis1995metis}.
%\lw{more info?}

Finally, because of the one-to-one mapping between a connectivity graph and a network graph, the vertex partitioning obtained on the connectivity graph can be easily translated back to the edge partitioning of the original graph. Algorithm~\ref{alg:TheAlgorithm} presents the key operations in Jigsaw algorithm.

%\textcolor{blue}{The optimisation problem above has been shown as NP-hard\cite{???}. We apply "what discrete optimisation technique" to find a nearly-optimal solution. "???" can achieve ... what is the gap???} Due to the equivalence of the two problems\cite{???}, the optimal solution for a connectivity graph can be easily translated back to the optimal edge partition for the original graph. Algorithm~\ref{alg:TheAlgorithm} presents the key operations in Jigsaw partitioning algorithm.

\begin{comment}
\begin{figure}[t]
\centering
\includegraphics[angle=90,origin=c,trim = 14mm 8mm 45mm 8mm, clip, width=0.65\columnwidth]{figures/1755_weighted_poppers.pdf}
\caption{Illustration of a partitioned network (AS 1755). Bolded vertices are popper nodes.}\label{fig:fancyIllustration}
\end{figure}
\end{comment}

\begin{figure}[t]
\centering
\includegraphics[trim = 5mm 5mm 5mm 5mm, clip, width=0.95\columnwidth]{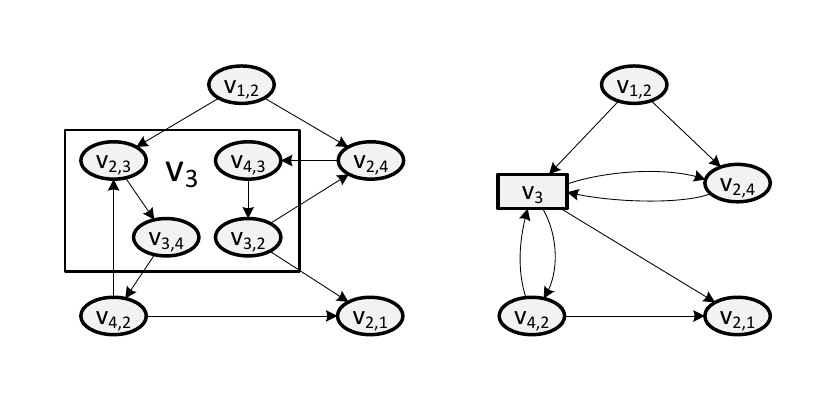}
%\caption{If $v_3$ cannot act as a popper switch, we can collapse the four connectivity graph vertices into one and still use the same partitioning algorithm.}\label{fig:connectivityGraphCollapsing}
\caption{Collapsing the connectivity-graph vertices adjacent to $v_3$ switch if it cannot act as a popper switch.}\label{fig:connectivityGraphCollapsing}
\end{figure}

\subsection{Partitioning heterogeneous networks}
\label{sec:hetero}

%%%\ma{Change ``legacy switch'' to something more sensible.}

%Connectivity graph brings significant flexibility when we partitioning a heterogeneous network wherein not all the switches can serve as a popper. 

So far, we have assumed that any switch can be selected as a popper. In some networks, however, only a subset of switches can perform the popping function while the rest, referred to as \emph{legacy switches}, can only act as forwarders. For example, regular OpenFlow switches cannot perform popping but can forward packets based on the iBF found in the IPv6-address fields~\cite{DBLP:ReedATTPS15}. 
% (e.g., OpenFlow switches can only be used as forwarders). 
We now briefly explain how the Jigsaw algorithm can be easily extended to handle these situations.

%Connectivity graph brings significant flexibility when we partitioning a heterogeneous network wherein not all the switches can serve as a popper due to many reasons (e.g., legacy routers). So far, we have assumed that any switch can be selected as a popper node. In some networks, however, only a subset of the switches can perform the popper function while the rest, called \emph{legacy switches}, can only act as forwarders. We now briefly explain how the Jigsaw-algorithm can be extended to these situations.

Recall that a switch is a popper iff it has incident links belonging to two or more different partitions. Thus, a legacy switch (i.e.\ forwarder) must have all of its links belonging to a single partition. This constraint can be fulfilled by collapsing all connectivity graph vertices together which represent links incident to a legacy switch. Figure~\ref{fig:connectivityGraphCollapsing} illustrates this in a scenario where $v_3$ is a legacy switch. All links incident to switch $v_3$ are aggregated to a single connectivity graph vertex.  In this way, we do not have to change the partitioning algorithm at all, but simply count vertex $v_3$ as four vertices during the partitioning phase of Jigsaw.
%However, the collapsing does not require us to change the partition algorithm at all except vertex $v_3$ must be count as four vertices during the partition.
%%%when the graph is finally being partitioned, the aggregated connectivity graph vertex $v_3$ must be count  as four vertices. This means that the used vertex partitioning algorithm must support non-uniform vertex-sizes.

Due to space constraints, the collapsing operation has been omitted in Algorithm~\ref{alg:TheAlgorithm}. Moreover, for the sake of clarity, in the rest of this paper we consider only networks where all switches can act as popper switches.

\section{Evaluation methodology}
\label{sec:evalMethod}

%Our evaluation is designed based on two distinct goals. First, we evaluate the partitioning quality of the Jigsaw-algorithm. We do this by comparing the Jigsaw to a popular edge partitioning heuristic. Second, we evaluate the scalability of the XBF forwarding scheme using various network topologies. More precisely, we increase the network size step step and compare the header overhead of XBF to that of the traditional Bloom-filter forwarding.

In this section, we explain our experiment setup and the evaluation metrics.

\newcommand{\scell}[2][l]{%
  \begin{tabular}[#1]{@{}l@{}}#2\end{tabular}}
\begin{table}\centering\scriptsize
\begin{tabular}{| p{2.0cm} | p{5.6cm} |}
\hline
\textbf{Algorithm} & \textbf{Parameters} \\\hline
Barab\'{a}si--Albert $G(n, m)$& $n \in \{500,1000,\smell,3000,4000,\smell8000\}$ \newline $m=2$  \\\hline
Erd\H{o}s--R\'{e}nyi $G(n, p)$ & $n \in \{500,1000,\smell,3000,4000,\smell8000\}$
\newline $p=(1 + \epsilon) \cdot n^{-1} \cdot \ln{n}$, $\epsilon=0.1$ 
\\\hline
\end{tabular}
%\caption{Summary of the synthetic topologies}
%\label{tbl:topologiesSummary}

~

~

\begin{tabular}{| p{2.0cm} | p{5.6cm} |}
\hline
\textbf{Partitioning algorithm} & \textbf{Parameters} \\\hline
Jigsaw & \scell{\#partitions $= 1.1 \cdot (|E|/256)$ \\ Vertex-partitioning with {\sc Metis}-library \cite{karypis1995metis} \\ {\sc Metis} parameters: ptype: k-way, iptype: grow, \\ ncuts: 1, objtype: vol} \\\hline
Powergraph~\cite{Gonzalez:2012:PDG:2387880.2387883} & \scell{target \# links in a partition: 256\\ algorithm applied recursively for partitions \\ with more links} \\\hline
\end{tabular}
\caption{Parameters used to create synthetic topologies and in partitioning algorithms}\label{tbl:parametersSummary}
\end{table}

\subsection{Experiment setup}
\label{sec:setup}

We use three different kinds of topologies in our evaluations: eight conventional ISPs, one community network, and various synthetic topologies. The ISP backbone topologies are based on the Rocketfuel $r_0$ datasets \cite{spring2002measuring}. We dropped the smallest ISP networks having less than 256 links. For the community network, we used the Guifi.net core network in Catalonia region~\cite{vega2012topology}. The synthetic topologies are generated using Barab\'{a}si--Albert (BA) and Erd\H{o}s--R\'{e}nyi (ER) models with various parameters (shown in Table~\ref{tbl:parametersSummary}). For all aforementioned topologies, we only consider the largest component.
%Most experiments are repeated 50 times to obtain representative results, the errorbars are omitted if the variance is small enough (i.e., 5\%).

\begin{figure}\centering
\includegraphics[trim = 0mm 0mm 0mm 0mm, clip, width=0.8\columnwidth]{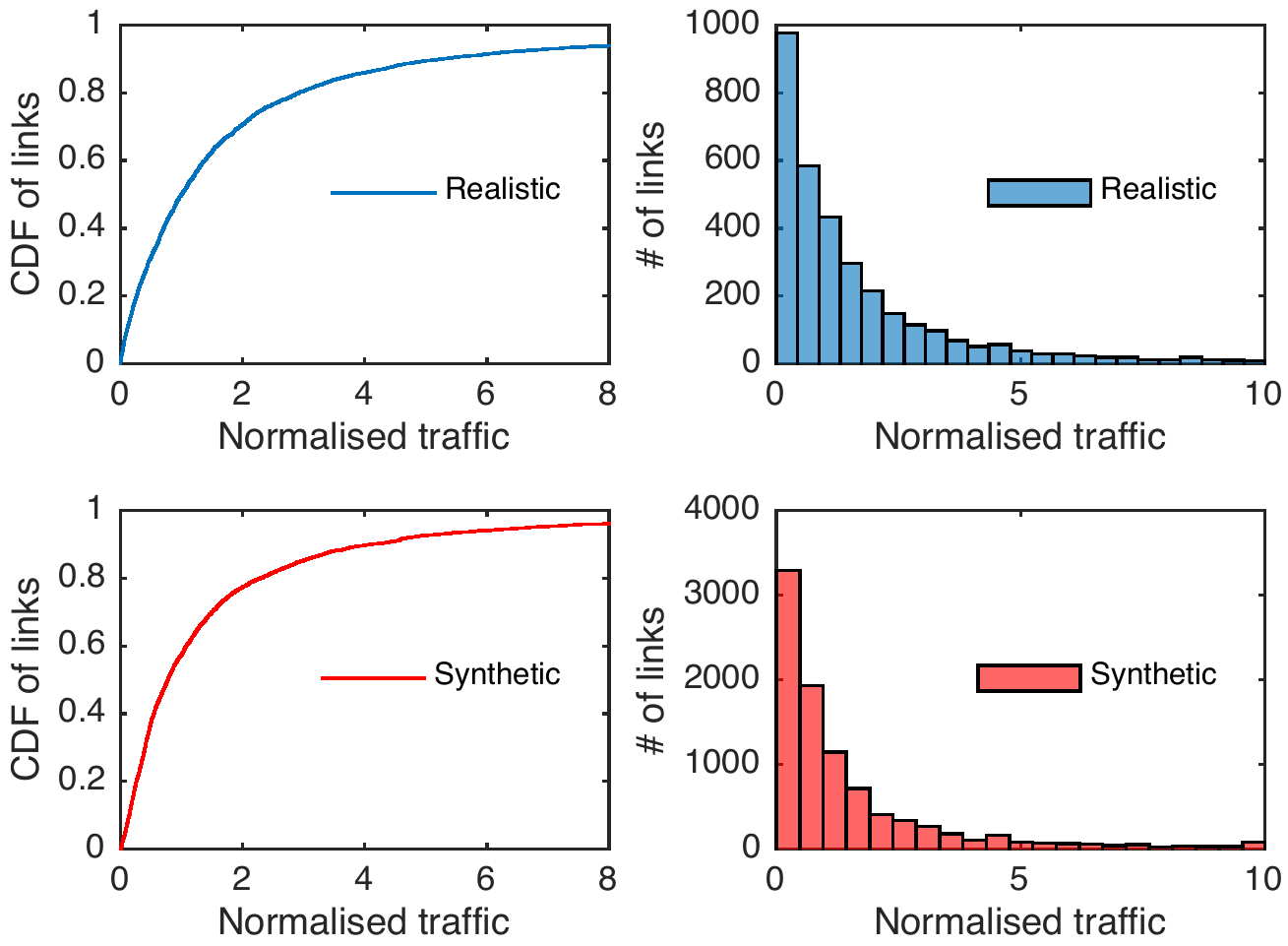}
\caption{Comparison of synthetic and realistic traffic patterns. Realistic traffic pattern was obtained with a  30-day monitoring of Guifi-network (Catalonia-region).}\label{fig:trafficPatterns}
\end{figure}

For traffic pattern, we use both realistic (for Guifi network) and synthetic traces (for ISP and synthetic networks). The realistic trace was obtained with a 30-day monitoring of Guifi network.
To generate synthetic traffic, we let every node communicate with any others with equal probability, namely we randomly select a certain number of data sinks for a given source with uniform distribution. 
A multicast tree is further constructed by combining the shortest paths between the data source and sinks.
Both realistic and synthetic traces contain the information of traffic volume on every directional link.
Figure~\ref{fig:trafficPatterns} shows the traffic distribution over all the links. Both trace types exhibit similar statistical distributions which justifies our traffic generation method.
We assume that the topology manager has \textit{a priori} knowledge about the traffic distributions in the network.

The vertex partitioning (i.e.\ the third step in our Jigsaw-algorithm, see Section~\ref{sec:partition}) was implemented with {\sc Metis}-library~\cite{karypis1995metis}, which we observed to produce high quality partitionings very fast (it took less than 20 seconds to partition the largest topologies with a mid-range laptop using one processor core). 
To evaluate the overall effectiveness of Jigsaw, we chose a popular edge partitioning heuristic called Powergraph~\cite{Gonzalez:2012:PDG:2387880.2387883} as our comparison baseline. The simulation was implemented with Python 2.7 using the Networkx, FNSS, pyMetis, and Scipy libraries. For each configuration, we repeated the experiments 1000 times to obtain statistically valid and representative results.

\subsection{Metrics}\label{sec:eval:metrics}

%%%The most obvious metric that we use to evaluate XBF is the total number of popper switches needed in a network. In addition to this, we calculate following metrics from the multicast trees obtained as described above.

In addition to the obvious metric like total number of poppers, we also calculate the following metrics to evaluate the effectiveness of XBF scheme.

\begin{itemize}
\item Mean number of partitions through which multicast trees traverse. This can be used to further calculate the header overhead of XBF.
\item Mean number of poppers through which multicast trees traverse, and mean number of popping operations for a single multicast packet. These can be used to evaluate the processing overhead in forwarding.
\end{itemize}

As mentioned, these metrics can be used to evaluate both the header overhead and the performance overhead of XBF. 
Moreover, to gain a comprehensive understanding of header overhead, for each topology, we calculate the minimum bit-lengths for Bloom-filters that would be required to store a multicast tree while keeping the number of false positives under a certain threshold. More specifically, we calculate values for function $\mathit{\textit{L}_{p}(s)}$, which tells the minimum bit-length for a Bloom-filter that can store a multicast tree with $s$ sinks in a given topology, while keeping $p$ percentage of the multicast trees free from false positives. Thus, $\mathit{\textit{L}_{99\%}(10)}$ denotes the length of a Bloom-filter that, given a topology, can store a multicast tree with 10 sinks while keeping 99\% of the multicast trees free from false positives.

%Moreover, to gain a comprehensive understanding of header overhead, for each topology, we calculate the minimum Bloom-filter length in order to guarantee that 50\% or 95\% of delivery trees will produce zero false positives. We refer to these metrics as  $\mathit{0\mbox{-}FP}_{50\%}$ and $\mathit{0\mbox{-}FP}_{95\%}$ respectively. 
%This metric is useful when comparing the header overheads of XBF and other Bloom-filter based forwarding schemes.

%In order to compare the header overheads with other Bloom-filter based forwarding solutions, we also calculated the expected bit-lengths for a Bloom-filter that does not cause any false-positive forwarding results. We calculated the expected bit-lengths for LIPSIN-type header and for Multi-stage Bloom-filter (MSBF) header, that are expected to cause zero false positives. The lengths were calculated using the equations 6--8 from~\cite{6877748}. The parameters required by the equations (i.e. delivery tree depth, number of links on a delivery tree, and number of links adjacent to the delivery tree) were obtained numerically by randomly generating 10000 delivery trees in each topology and by averaging the results.

\section{Evaluation results}
\label{sec:evalResults}

The most important evaluation results have been compiled into three groups (i.e., unicast with just one sink, multicast with ten sinks and with 20 sinks) in Table~\ref{tlb:bigAssTable} along with some statistics about the topologies. 
%%%We now go these results through in detail. Section~\ref{sec:eval:summary} summarises the evaluation results. 
Throughout this section, the whisker bars in the box-plots represent the 5th and 95th percentile values of the measurement while the boxes show the interquartile range. 

%We start going through these results by first analysing the partitioning quality of the Jigsaw algorithm. After this, we analyse the overall scalability of the XBF forwarding scheme in terms of growing multicast trees and growing networks. 

\begin{figure*}[t]
\minipage[t]{0.24\textwidth}\raggedleft
\includegraphics[trim = 0mm 0mm 0mm 5mm, clip, width=0.96\columnwidth]{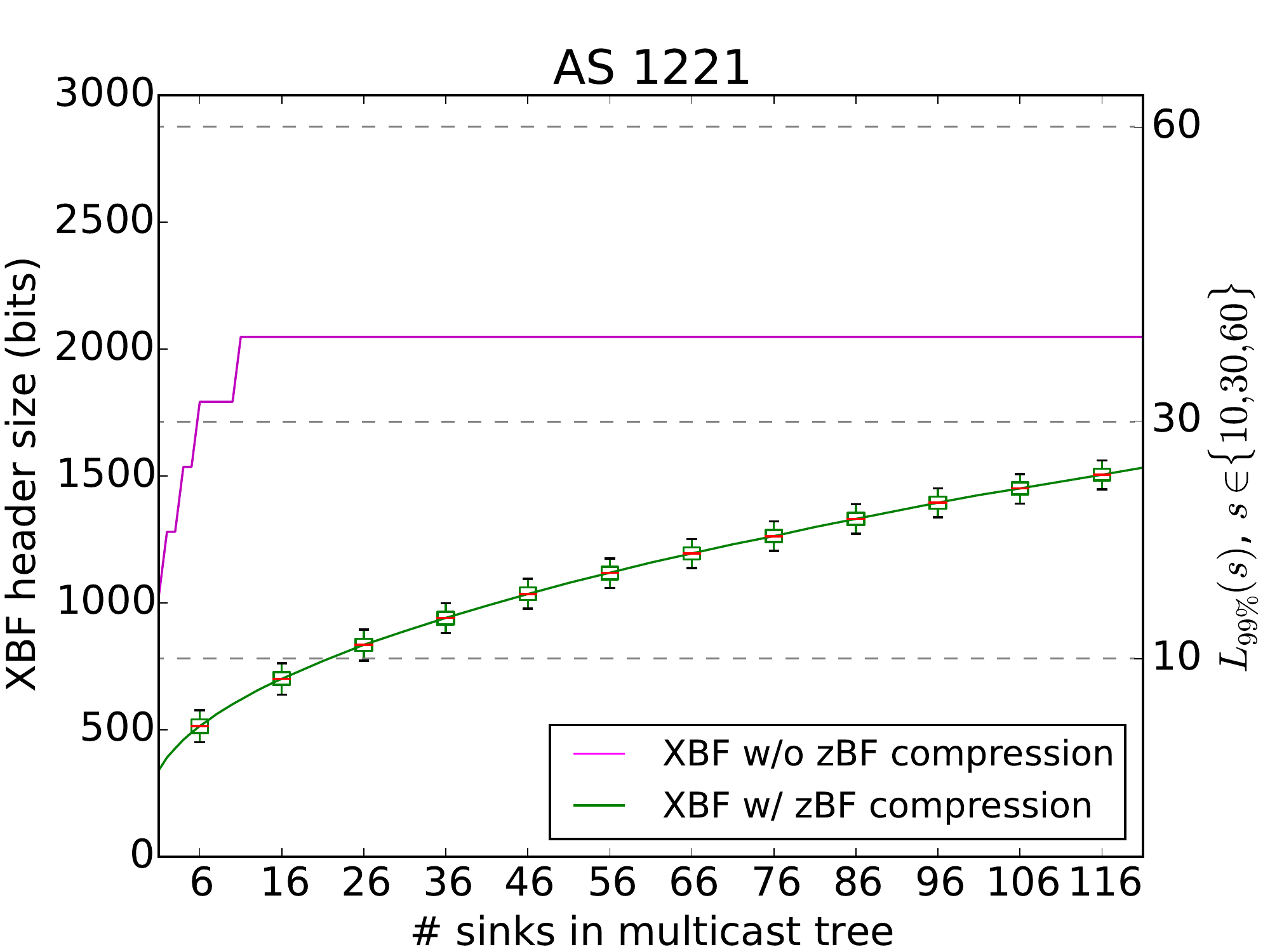}
\caption{XBF header overhead. For the horizontal lines, see text}\label{fig:headerOverhead}
\endminipage\hfill
\minipage[t]{0.24\textwidth}\raggedleft
	\includegraphics[trim = 0mm 0mm 0mm 5mm, clip, width=0.96\textwidth]{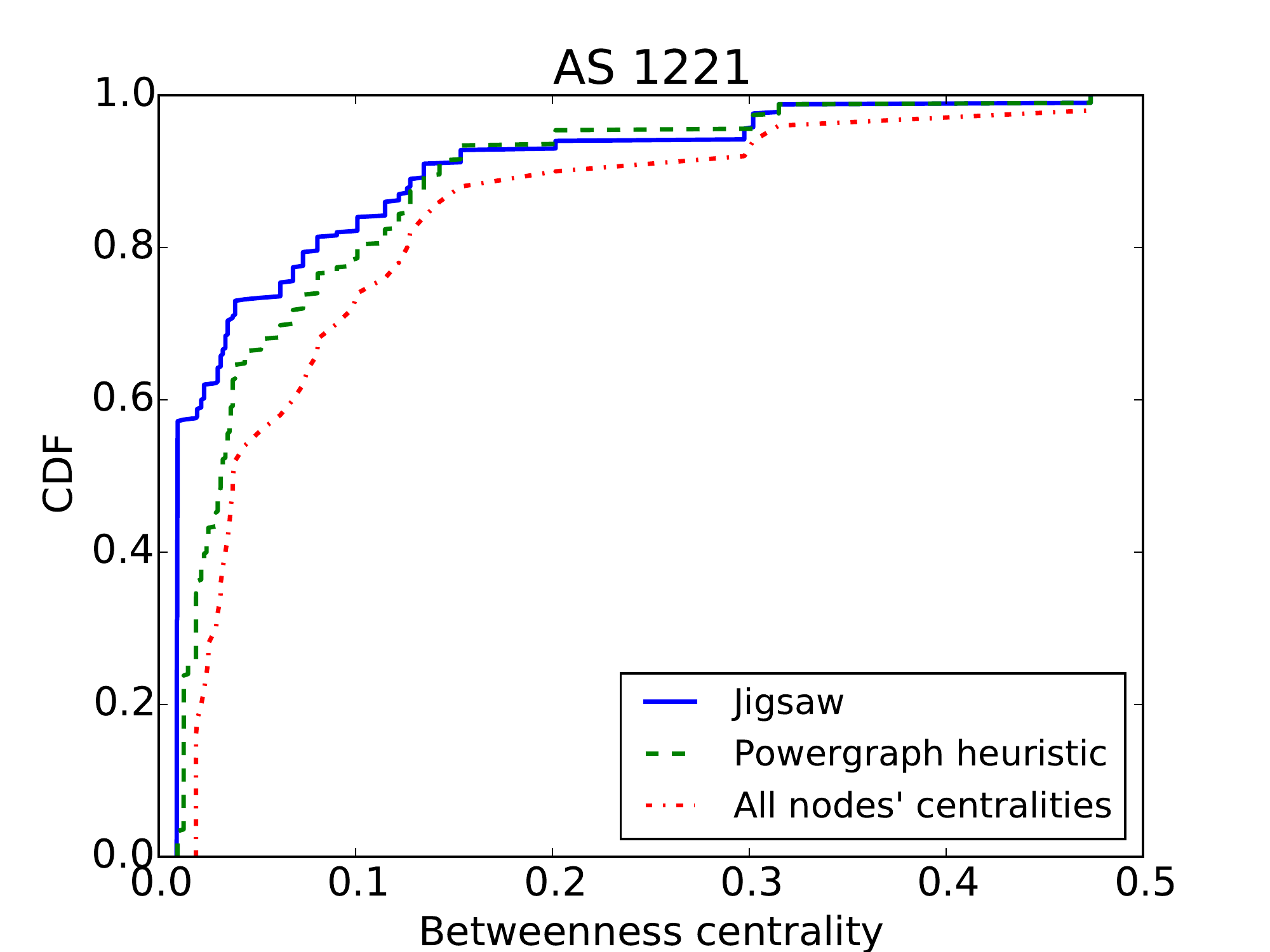}
	\caption{Betweenness centralities of the 50 most central popper nodes}\label{fig:betweennessCentsCDF}
\endminipage\hfill
\minipage[t]{0.48\textwidth}\raggedleft
	\subfigure[Poppers on a multicast tree as function of no.\ sinks]{%
	\includegraphics[trim = 0mm 0mm 5mm 5mm, clip, width=0.47\textwidth]{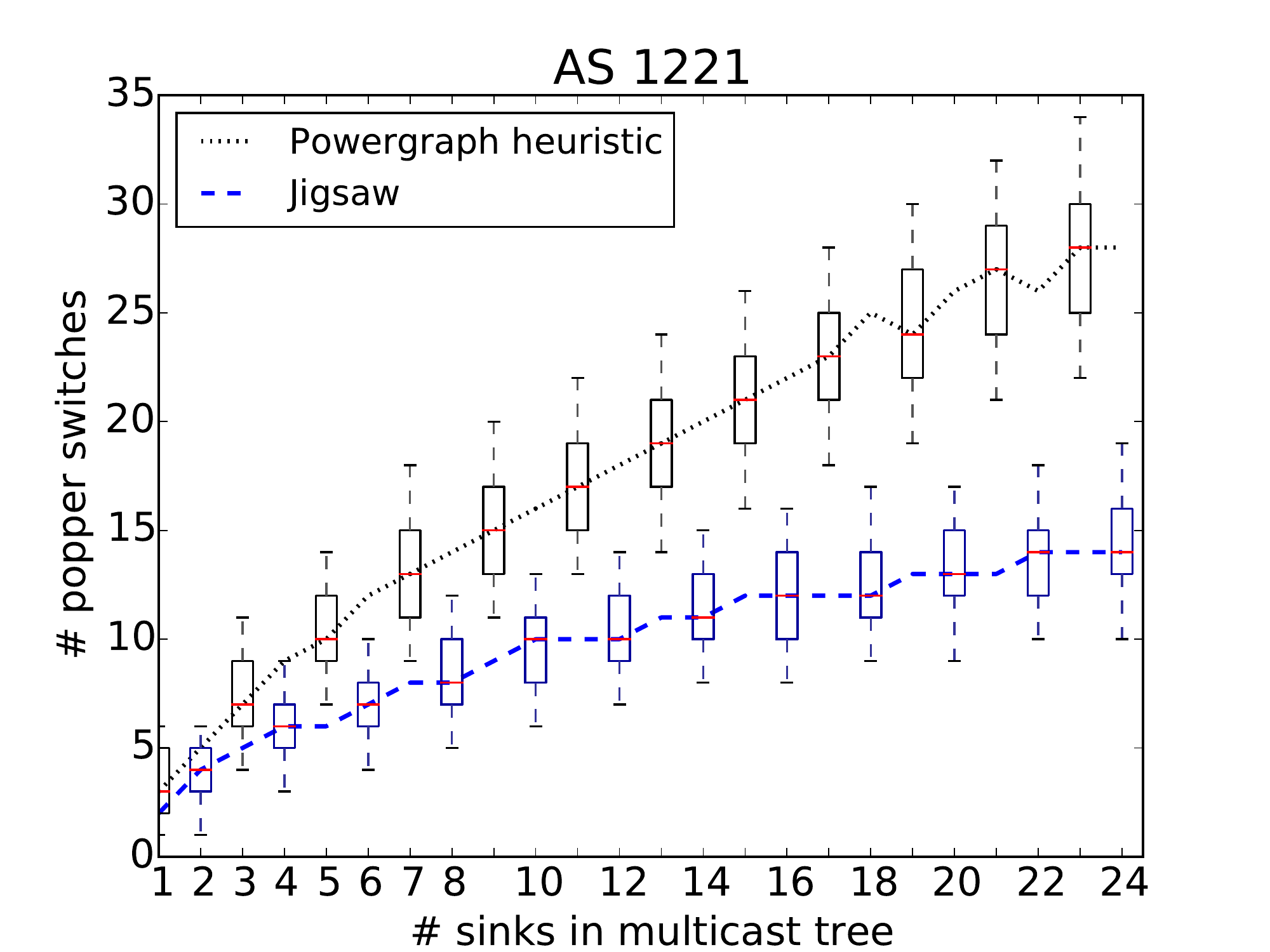}%
	\label{fig:noPopperSwitches}}\quad%
	\subfigure[No.\ popping operations as function of no.\ sinks]{%
	\includegraphics[trim = 0mm 0mm 5mm 5mm, clip, width=0.47\textwidth]{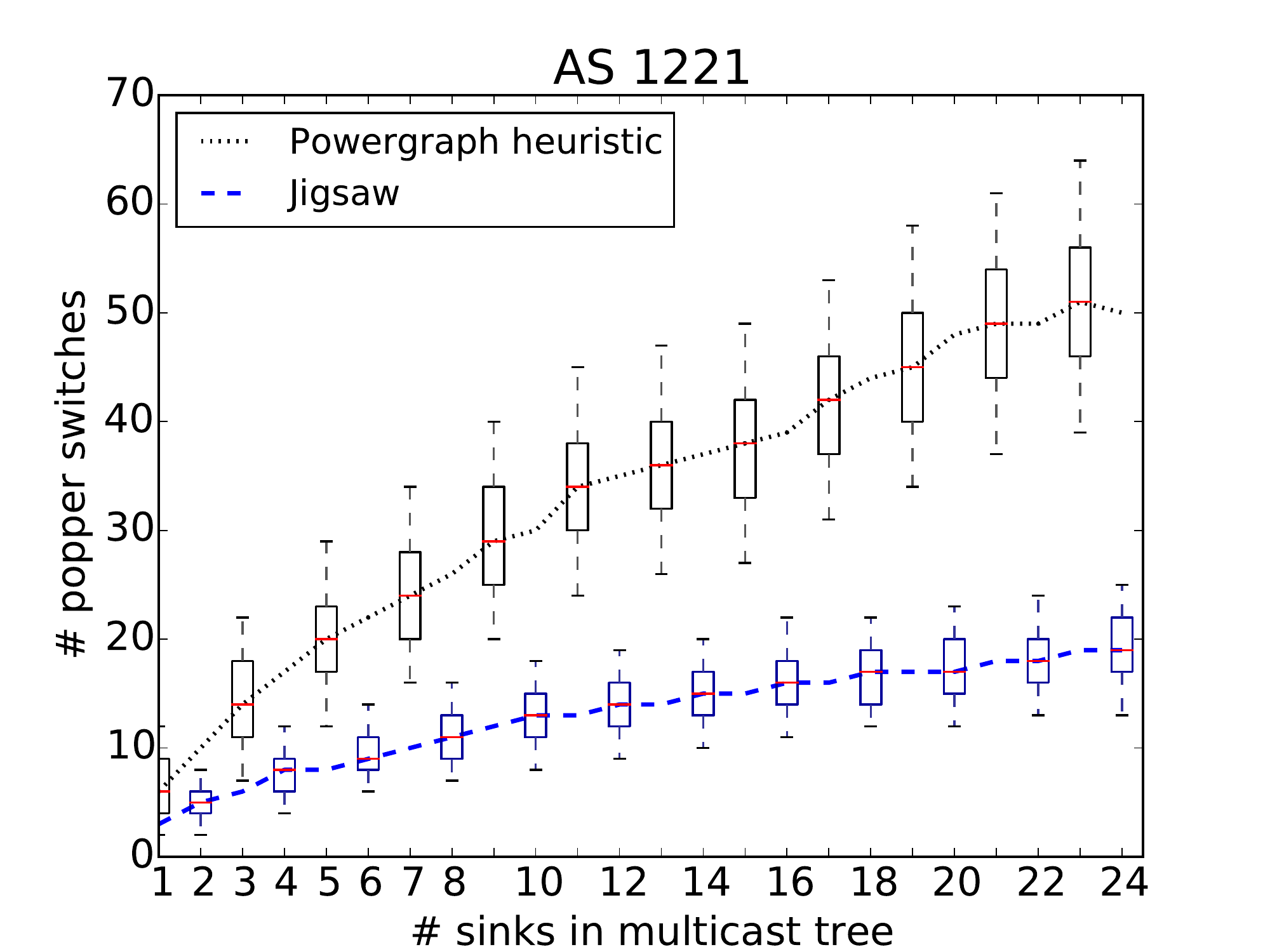}%
	\label{fig:noPoppings}}\vspace{-0.5\baselineskip}
	\caption{Comparison of partitioning algorithms}
\endminipage\hfill
\end{figure*}

\begin{comment}
\begin{figure*}[t]
\minipage[t]{0.3\textwidth}
\includegraphics[trim = 0mm 0mm 5mm 5mm, clip, width=1\columnwidth]{figures/scalability_poppers.pdf}
\caption{No.\ popper swiches through which a multicast tree (10 sinks) passes as a function of network size}\label{fig:scalabilityPoppers}
\endminipage\hfill
\minipage[t]{0.3\textwidth}
\includegraphics[trim = 0mm 0mm 5mm 5mm, clip, width=1\columnwidth]{figures/scalability_poppings.pdf}
\caption{No.\ popping operations that occur during a multicast transmission (10 sinks) as a function of network size)}\label{fig:scalabilityPoppings}
\endminipage\hfill
\minipage[t]{0.3\textwidth}
\includegraphics[trim = 0mm 0mm 5mm 5mm, clip, width=1\columnwidth]{figures/scalability_header.pdf}
\caption{XBF header overhead (with compression) and $L_{95\%}(10)$ as a function of network size (multicast tree with 10 sinks)}\label{fig:scalabilityHeaderOverhead}
\endminipage\hfill
\end{figure*}
\end{comment}

\begin{figure*}[t]
\centering
\subfigure[No.\ popper switches through which a multicast tree (10 sinks) passes]{\includegraphics[trim = 0mm 0mm 15mm 5mm, clip, width=0.31\textwidth]{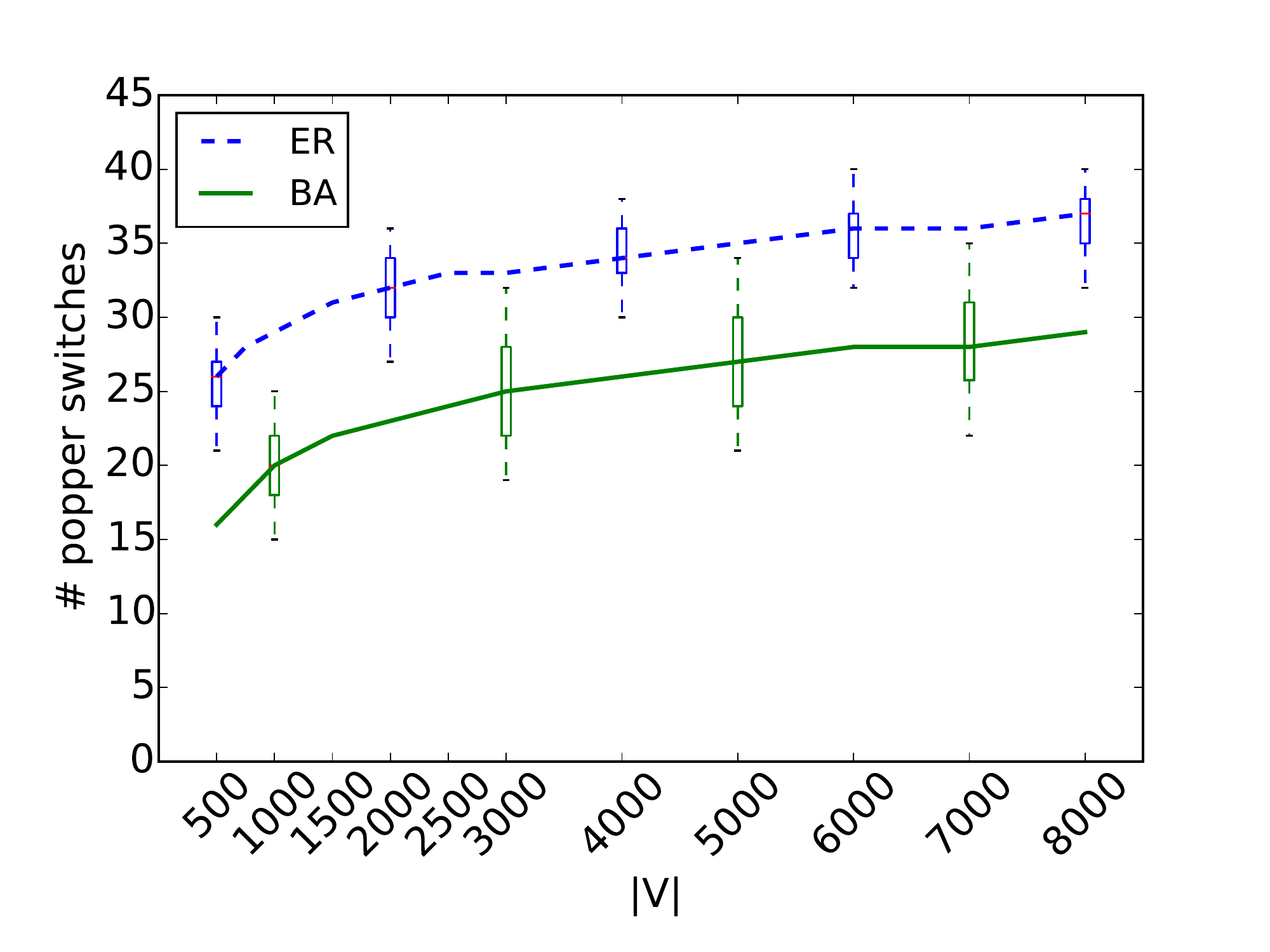}\label{fig:scalabilityPoppers}}\hfill
\subfigure[No.\ popping operations that occur during a multicast transmission]{
\includegraphics[trim = 0mm 0mm 15mm 5mm, clip, width=0.31\textwidth]{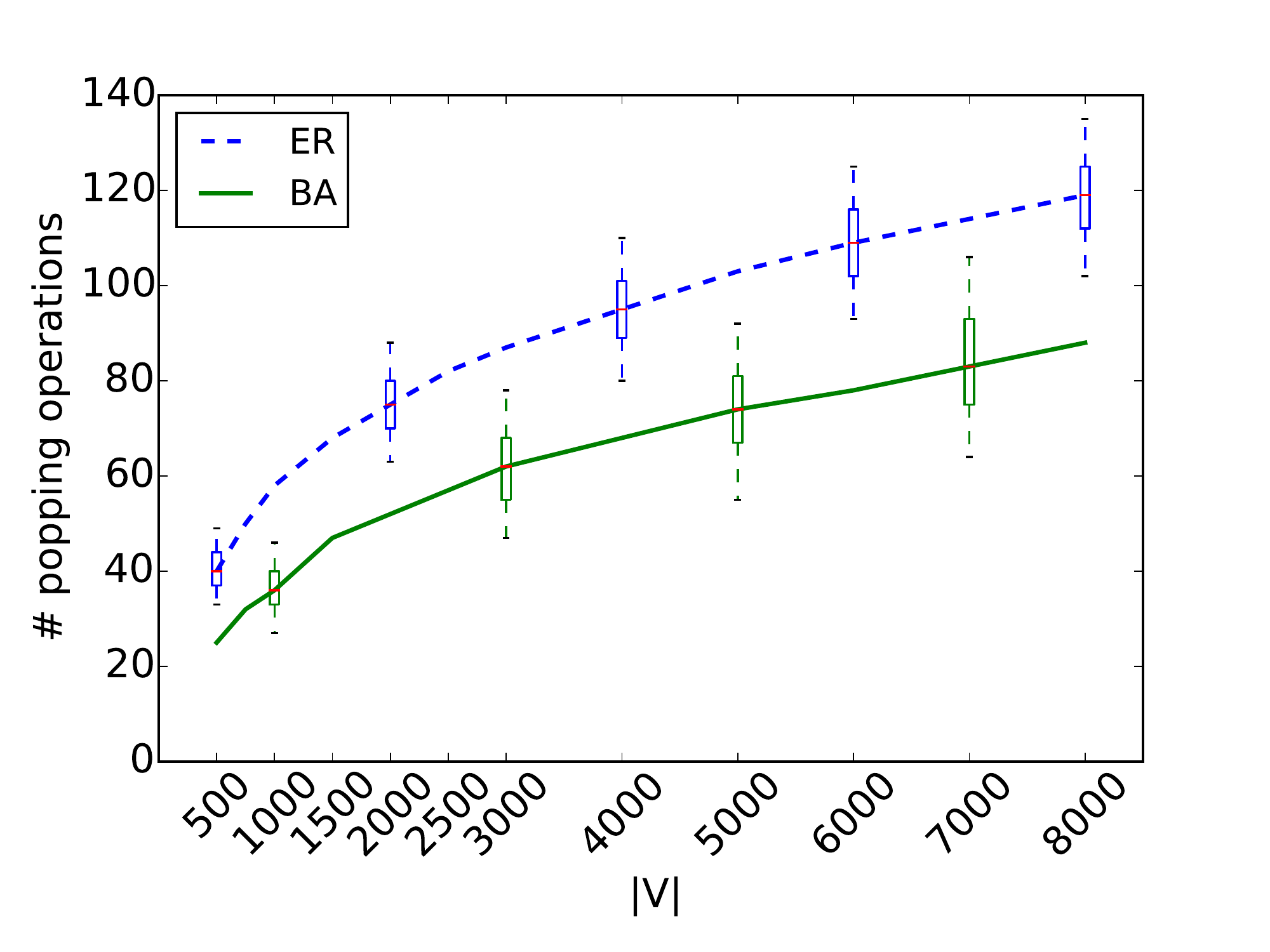}\label{fig:scalabilityPoppings}}\hfill
\subfigure[XBF header overhead (with compression) and $L_{95\%}(10)$]{
\includegraphics[trim = 0mm 0mm 15mm 5mm, clip, width=0.31\textwidth]{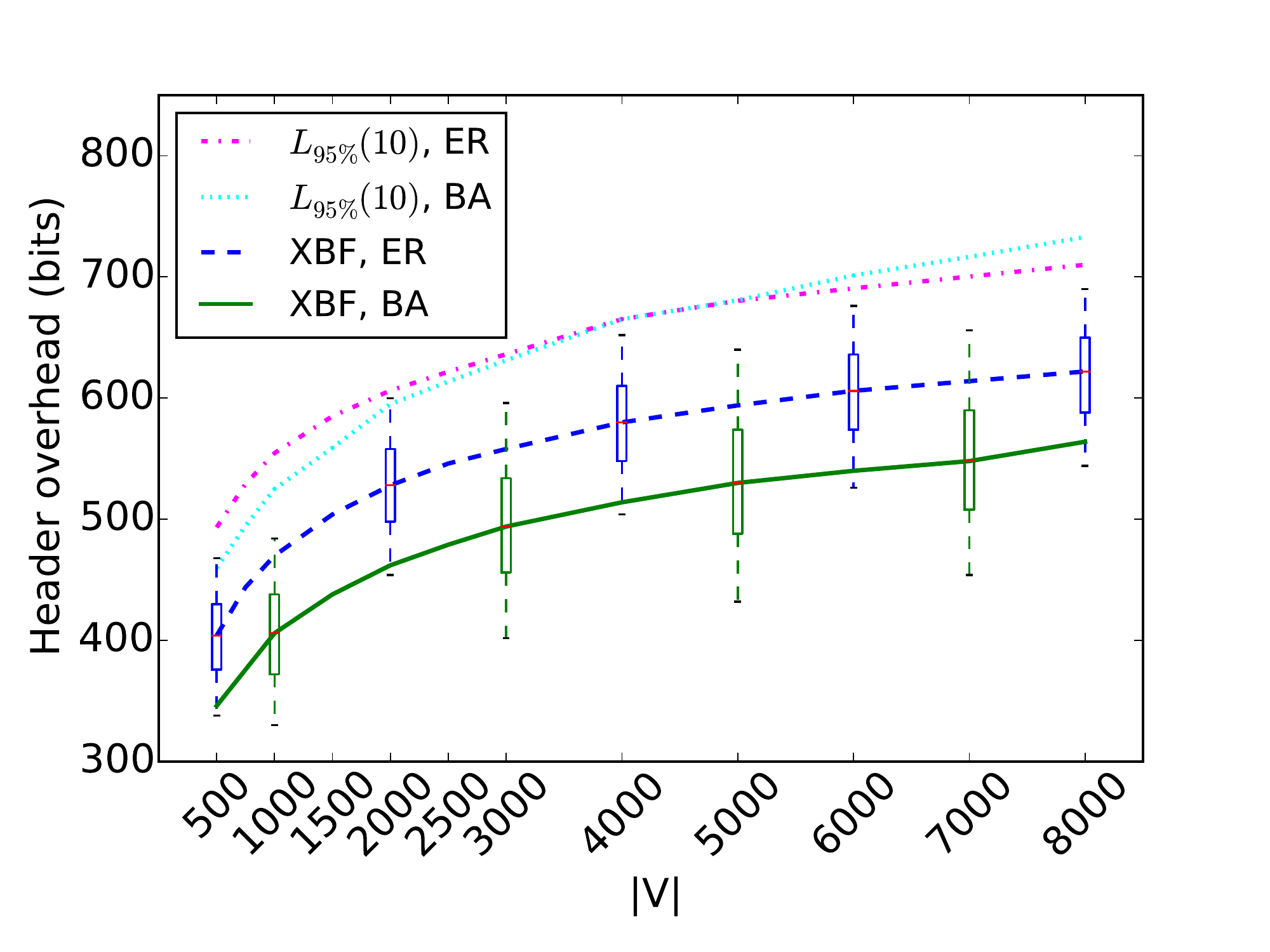}\label{fig:scalabilityHeaderOverhead}}\vspace{-0.5\baselineskip}
\caption{Different metrics calculated from a multicast tree with 10 sinks, as a function of topology size (number of nodes). Some error bars omitted for clarity.}
\end{figure*}

\newcommand{\thead}[1]{\centering \textbf{#1}}
\newcommand{\trhead}[1]{\rotatebox{270}{\parbox{2.8cm}{\thead{#1}}}}

\begin{table*}[t]\centering\scriptsize
\begin{tabular}{| l c || c | c | c | c | c | c || c | c | c | c || c | c | c | c || c | c | c | c |}
\cline{9-20}
\multicolumn{8}{c||}{}  & \multicolumn{4}{|c||}{\thead{One sink (unicast)}} & \multicolumn{4}{|c||}{\thead{Ten sinks (multicast)}} & \multicolumn{4}{|c|}{\thead{20 sinks (multicast)}} \\\hline

\multicolumn{2}{|c||}{\trhead{Topology}} & \trhead{\# nodes} & \trhead{\# directed links} & \trhead{Mean path len.\ (nodes)} & 
\trhead{\# partitions} & \trhead{\# popper switches} & \trhead{Mean \# popping switches on path} & 

\trhead{Mean header overhead} & \trhead{Mean hdr.\ overhead (compressed zBF)} & 
\trhead{$L_{95\%}(1)$} & \trhead{$L_{99\%}(1)$} & 

\trhead{Mean header overhead} & \trhead{Mean hdr.\ overhead (compressed zBF)} & 
\trhead{$L_{95\%}(10)$} & \trhead{$L_{99\%}(10)$} &

\trhead{Mean header overhead} & \trhead{Mean hdr.\ overhead (compressed zBF)} & 
\trhead{$L_{95\%}(20)$} & \trhead{$L_{99\%}(20)$} 
\\ 

%\hline%\hline
\hhline{==#=|=|=|=|=|=#=|=|=|=#=|=|=|=#=|=|=|=|}

\multirow{8}{*}{\rotatebox{270}{Rocketfuel AS}}
& 1755 & 172 & 762 & 5.9 & 4 & 19 & 1.5 & 758 & 338 & 83 & 103 & 1251 & 583 & 514 & 617 & 1279 & 725 & 849 & 1013 \\
& 3257 & 240 & 808 & 6.6 & 4 & 35 & 1.4 & 772 & 346 & 90 & 112 & 1244 & 583 & 496 & 595 & 1276 & 726 & 837 & 999 \\
& 3967 & 201 & 868 & 5.8 & 4 & 21 & 1.4 & 762 & 337 & 79 & 98 & 1256 & 592 & 519 & 622 & 1279 & 733 & 868 & 1302 \\
& 1221 & 318 & 1516 & 6.0 & 7 & 72 & 2.3 & 945 & 346 & 96 & 117 & 1870 & 602 & 510 & 765 & 2013 & 761 & 847 & 1270 \\
& 7018 & 631 & 4156 & 6.0 & 18 & 191 & 2.9 & 992 & 347 & 90 & 111 & 2884 & 706 & 650 & 975 & 3691 & 953 & 1128 & 1693 \\
& 1239 & 604 & 4536 & 5.2 & 20 & 310 & 3.1 & 982 & 336 & 79 & 96 & 3168 & 683 & 573 & 860 & 4133 & 939 & 1031 & 1546 \\
& 2914 & 960 & 5642 & 6.2 & 25 & 308 & 3.4 & 1078 & 352 & 96 & 117 & 3613 & 762 & 690 & 1035 & 4861 & 1079 & 1242 & 1864 \\
& 3356 & 624 & 10596 & 4.4 & 46 & 461 & 2.8 & 929 & 325 & 73 & 88 & 3865 & 655 & 535 & 803 & 5768 & 933 & 1000 & 1501 \\
\hline
\multicolumn{2}{|l||}{\multirow{3}{*}{\parbox{1.2cm}{Barab\'asi--Albert}}}
& 500 & 1992 & 4.7 & 9 & 287 & 2.8 & 875 & 328 & 70 & 85 & 2173 & 599 & 459 & 689 & 2453 & 793 & 817 & 1225 \\
\multicolumn{2}{|l||}{}
& 1000 & 3992 & 5.0 & 18 & 646 & 3.1 & 961 & 334 & 79 & 96 & 3107 & 662 & 519 & 779 & 3994 & 916 & 948 & 1422 \\
\multicolumn{2}{|l||}{}
& 2000 & 7992 & 5.3 & 35 & 1432 & 3.6 & 1031 & 340 & 87 & 105 & 4020 & 722 & 592 & 887 & 5739 & 1035 & 1076 & 1615 \\
\hline
\multicolumn{2}{|l||}{\multirow{3}{*}{\parbox{1.2cm}{Erd\H{o}s--R\'enyi}}}
& 497 & 3496 & 4.4 & 16 & 468 & 3.1 & 854 & 324 & 57 & 71 & 2875 & 660 & 491 & 588 & 3661 & 909 & 930 & 1395 \\
\multicolumn{2}{|l||}{}
& 999 & 7732 & 4.6 & 34 & 979 & 3.5 & 931 & 328 & 62 & 77 & 3956 & 724 & 552 & 658 & 5648 & 1044 & 1051 & 1577 \\
\multicolumn{2}{|l||}{}
& 2000 & 16730 & 4.8 & 72 & 1970 & 3.8 & 990 & 333 & 66 & 83 & 4883 & 783 & 607 & 910 & 7649 & 1173 & 1174 & 1760 \\
\hline
\multicolumn{2}{|l||}{\parbox{1.2cm}{Guifi, Catalonia}}
& 7168 & 14872 & 11.7 & 64 & 1145 & 5.3 & 1617 & 433 & 222 & 333 & 5779 & 1042 & 1190 & 1785 & 8227 & 1468 & 1946 & 2919 \\\hline

\end{tabular}
\caption{Summary of the evaluation results on both realistic and synthetic topologies using different number of sinks.}\label{tlb:bigAssTable}
\end{table*}

\subsection{Header overhead}
\label{sec:eval:headerOverhead}

We start by analysing the header size in XBF. Recall from Section~\ref{sec:forwarding} that the header overhead in XBF is the aggregated size of iBF and zBF. 
%The size of the iBF is always 256 bits, while the length of the zBF varies depending on the number of filters in it, and how well it can be compressed. \lw{is here the first time we mention compressing?}
The size of the iBF is always 256 bits, while the length of the zBF varies depending on the number of filters in it, which again depends on the number of partitions a multicast tree traverses through. Also, since the zBF is accessed only by relatively few nodes (that is, the popper switches), it may be compressed.
Here, we present the header overheads with and without such compression. The compression was done simply by compressing the entire zBF with with binary run length encoding (we use \textit{Elias-gamma coding} for the run lengths~\cite{elias1975universal}). %Although the compression rate degrades as more and more links are included, this simple coding can achieve up to 70\% saving on header overhead (see Fig.~\ref{fig:headerOverhead}).

%The number of filters in zBF equals the number of partitions the multicast tree traverses through (except when the tree passes through only one partition in which case the zBF can be removed altogether). 
%Thus, if we let $\rho_t$ denote the number of partitions a multicast tree $t$ passes through, the header overhead is ${256 \times \rho_t + 256}$ bits for ${\rho \geq 2}$.

%The ?th column in Table~\ref{tlb:bigAssTable} shows the mean number of popper-switches that a unicast path crosses for every topology. Depending on the topology, $\bar{\rho}$ varies between 1.4--5.8. The small variation of $\bar{\rho}$ can be well explained by the fact that most networks have relatively small diameters in practice (figures omitted due to space constraints). As we will see later, this property contributes positively to the performance of XBF. \ma{FIX THIS!}

Table~\ref{tlb:bigAssTable} shows the mean header overheads in all topologies when the multicast tree has 1 (unicast), 10, or 20 sinks. 
%The table also shows the mean header overheads when the zBF is being compressed. 
It can be seen that the header overhead is largely a function of the multicast tree size -- although the topology also affects it, its effect is relatively moderate. This is further illustrated in
%The XBF header overhead is largely a function of the multicast tree size. This is illustrated in
Figure~\ref{fig:headerOverhead} which plots the header overhead as a function of different tree sizes in AS 1221 topology\footnote{There is no particular reason why we created the figures using the data from AS 1221.}. The horizontal dashed lines indicate the optimal bit-length for a Bloom-filters that can store a multicast tree with $s$ sinks while keeping 99\% of multicast trees false-positive free. More formally, they show the value $L_{99\%}(s)$ for $s \in \{10,30,60\}$. %\lw{Did you define $L_{99\%}(s)$ anywhere?}
%The figure also shows what would be the length of an optimally filled regular Bloom-filter that could store the given sized multicast tree without producing any false positives (or more informally, what is the length of LIPSIN-style Bloom-filter that causes no false positives~\cite{Jokela:2009}). 

It can be seen from the figure that when a multicast trees grows, the size of the non-com\-pressed XBF header increases accordingly since more partitions are involved in a distribution. However, zBF can be easily compressed due to its sparsity. In other words, using a single bit to identify one link will create multicast tree representations mostly filled with zeroes. (This is not necessarily true for traditional LIPSIN-like Bloom-filters forwarding where several pseudo randomly selected bits are set to one for every link that is included in the multicast). % tree therefore the increasing the entropy of the filter and making compression very difficult.
%roughly half of the bits are randomly set to 1 and are uniformly distributed in the string (due to the hash functions), therefore the induced high entropy makes compressing very difficult.
%%%n fact, an optimally filled Bloom-filter cannot be compressed at all because it has 50\% of the randomly selected bits set to one~\cite{?}.
Thus, with the simple run-length encoding done on the the zBF, it is possible to achieve up to 70\% saving on header overhead.

%We compress every individual Bloom-filter in the zBF with binary run length encoding (we use \textit{Elias-gamma coding} for the run lengths~\cite{elias1975universal}). Although the compression rate degrades as more and more links are included, this simple coding can achieve up to 70\% saving on header overhead (see Fig.~\ref{fig:headerOverhead}).

%So far, we have focused on comparing the Jigsaw algorithm to our baseline graph partitioning approach. We now move onwards and consider how XBF performs in terms of header size and compare it to the bit-length of a Bloom-filter that on average does not produce any false positives. 

%It is clear from the values shown in Tables~\ref{tlb:bigAssTable}--\ref{tbl:secondBigAssTable} that when the multicast tree is small, XBF will produce has larger headers compared to LIPSIN-style Bloom-filter\footnote{The header overhead of XBF header is 256 times the number of partitions in the multicast tree.}. With unicast, the header overhead of XBF is depending on topology on average between 512--724 bits. With LIPSIN, encoding the path into a 70--120 bit long Bloom-filters produces false positives only with 5\% probability. While the difference is large, it diminishes when the more receivers are added to the multicast tree. With ten receivers, the difference has lowered to 1000--3800 bits (XBF) versus 630--880 bits (LIPSIN). Figure~\ref{fig:headerOverhead} shows the header overheads with even larger delivery trees using the AS 1221 topology as an example. 

\subsection{Processing overhead in forwarding}
\label{sec:eval:processingOverhead}

The key goal of XBF is to allow false-positive free forwarding while simultaneously minimizing the processing overhead of the packet forwarding. While basic forwarder switches can forward packets with minimal overhead, the same is not true for popper switches. We now evaluate the \emph{processing overhead} that is caused by the inter-partition traffic and popping operations. The key metrics used here are (1) the number of popper switches in the network, (2) the number of popper switches on a multicast tree, and (3) the number of popping operations occurring during packet forwarding. %%%While these values correlate, they are not the same.

As already discussed, the processing overhead is dependent on how the network is partitioned. Thus, the performance of Jigsaw algorithm plays a central role in the performance of XBF. We now compare Jigsaw algorithm to another relatively simple edge-partitioning heuristic called Powergraph. The purpose of this comparison is to validate the performance of Jigsaw and to show the importance of the partitioning algorithm. 

We observed that our Jigsaw-algorithm outperforms the Powergraph-heuristic consistently in all topologies.
%
%Our results show that Jigsaw-algorithm outperforms the Powergraph-heuristic consistently in all topologies. The performance difference is smaller in the topologies generated with the Barab\'asi--Albert preferential-attachment model. %%%While this is hardly surprising, it validates our hypothesis regarding the importance of a good partitioning algorithm.
%
A notable difference between the performances of Jigsaw and Powergraph heuristic is that the partitioning given by the Powergraph heuristic requires much more popping switches. Furthermore, Powergraph-heuristic places the popper switches to more central locations (i.e., network core) compared to the Jigsaw-algorithm. This can be seen in Figure~\ref{fig:betweennessCentsCDF} which shows the betweenness centrality values of the most central popper nodes -- the popper switches' centrality values are much higher when the partitioning is done with Powergraph heuristic compared to the partitioning done by Jigsaw. This basically means that the Powergraph heuristic not only produces much more popping nodes, but also places them to worse places in the network.

%the betweenness centrality values of the most central popper nodes are much higher when the partitioning is done with Powergraph heuristic compared to the partitioning done by Jigsaw. This basically means that the Powergraph heuristic not only produces much more popping nodes, but also places them to worse places in the network.
%A notable difference between the performances of Jigsaw and Powergraph heuristic is that the partitioning given by the Powergraph heuristic requires much more popping switches. Furthermore the popping switches are located in more central locations (i.e., network core) compared to the ones assigned by the Jigsaw-algorithm. This can be seen in Figure~\ref{fig:betweennessCentsCDF}: the betweenness centrality values of the most central popper nodes are much higher when the partitioning is done with Powergraph heuristic compared to the partitioning done by Jigsaw. This basically means that the Powergraph heuristic not only produces much more popping nodes, but also places them to worse places in the network.

The placement of the poppers directly affects the number of popping operations that must be performed in packet forwarding. As a result, it is not surprising that the number of popping operations is also smaller when the network is partitioned using Jigsaw. This can be seen in figures~\ref{fig:noPopperSwitches}--\ref{fig:noPoppings} which show the number of poppers on a multicast trees as well as the number of popping operations that happen during forwarding. Furthermore, the figures show that the consequences of a poor partitioning are emphasized when the multicast trees grow.

\begin{figure*}%
\minipage[t]{0.32\textwidth}	
\includegraphics[trim = 0mm 0mm 0mm 0mm, clip, width=1\textwidth]{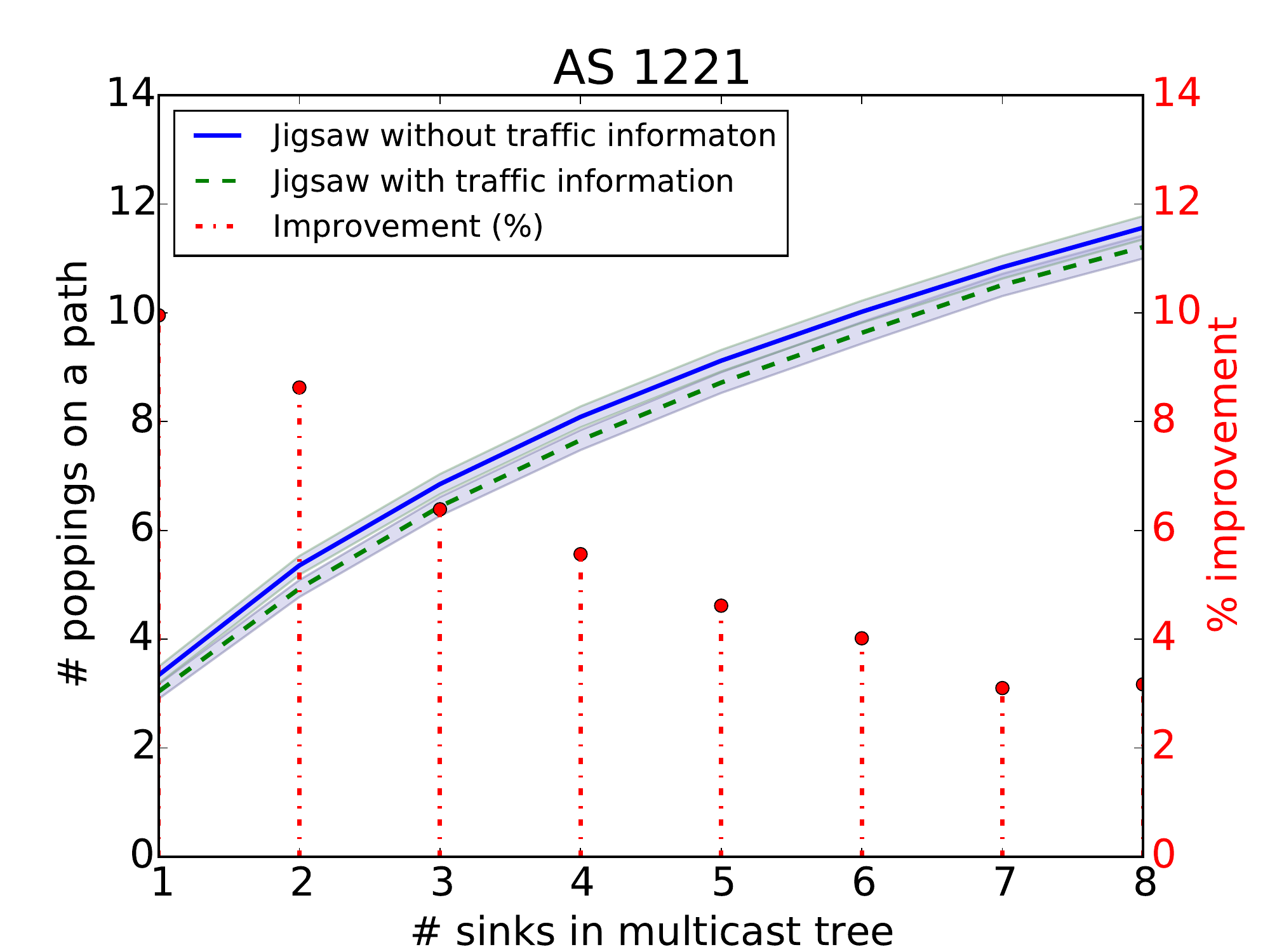}
\caption{Partitioning quality improvement when traffic patterns are taken into account (details in text)}\label{fig:poppingsSkewed}
\endminipage\hfill
\minipage[t]{0.64\textwidth}
\centering
\subfigure{\includegraphics[angle=90,origin=c,trim = 14mm 8mm 45mm 8mm, clip, width=0.40\textwidth]{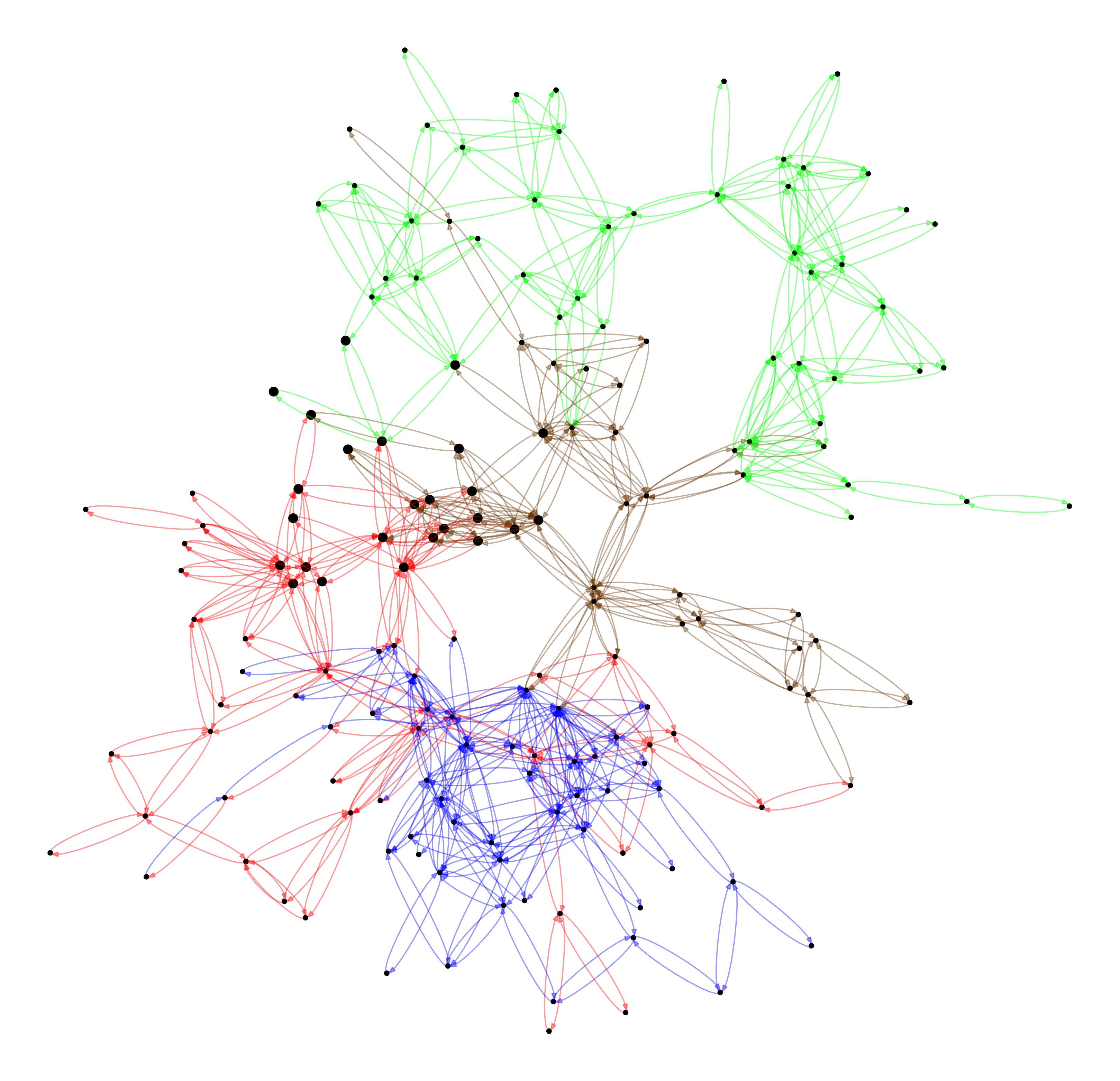}
\llap{\raisebox{0cm}{
      \includegraphics[angle=90,origin=c,trim = 14mm 8mm 45mm 8mm, clip,width=0.40\textwidth]{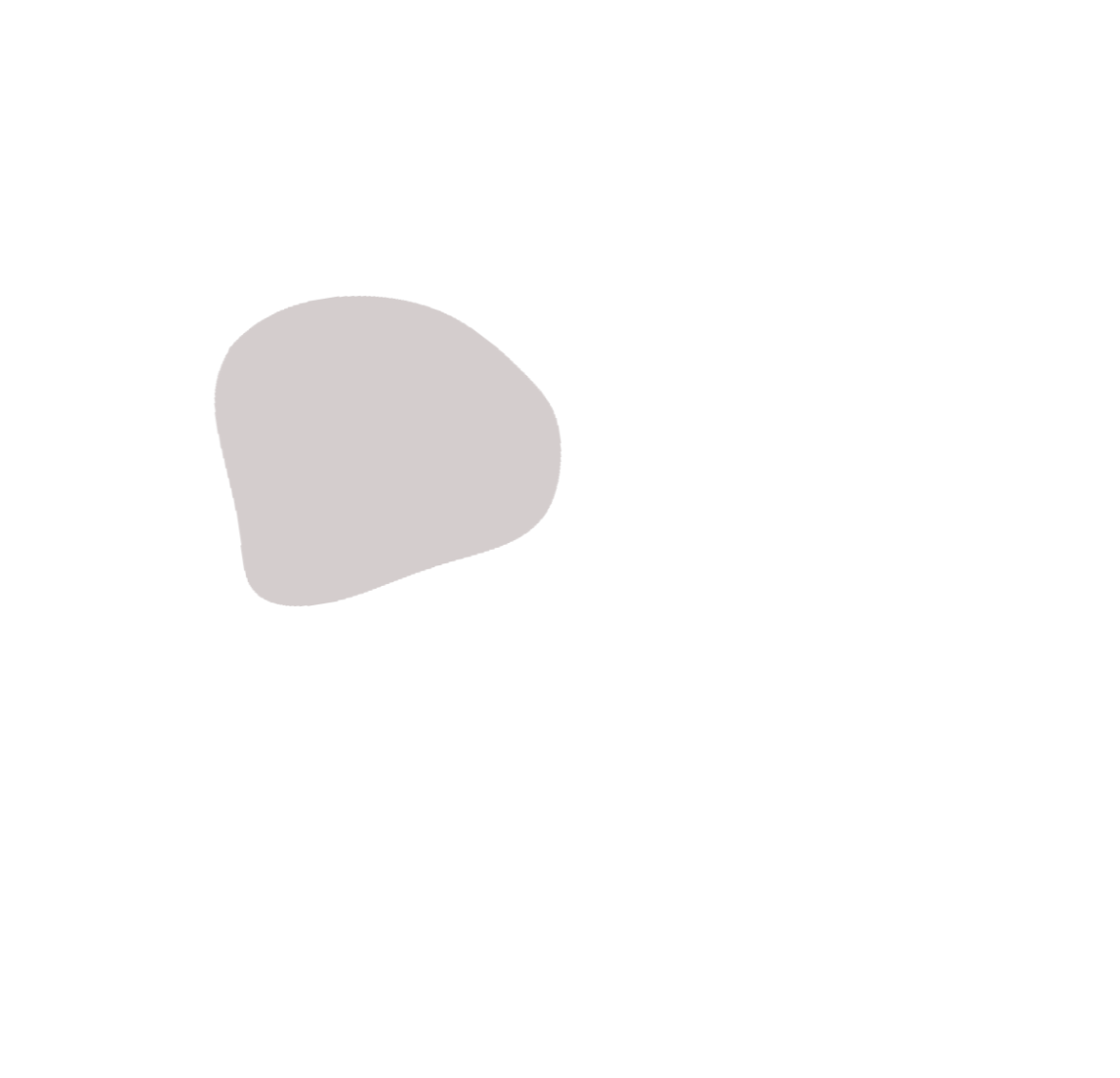}%
    }}%
}\qquad
\subfigure{\includegraphics[angle=90,origin=c,trim =14mm 8mm 45mm 8mm, clip, width=0.40\textwidth]{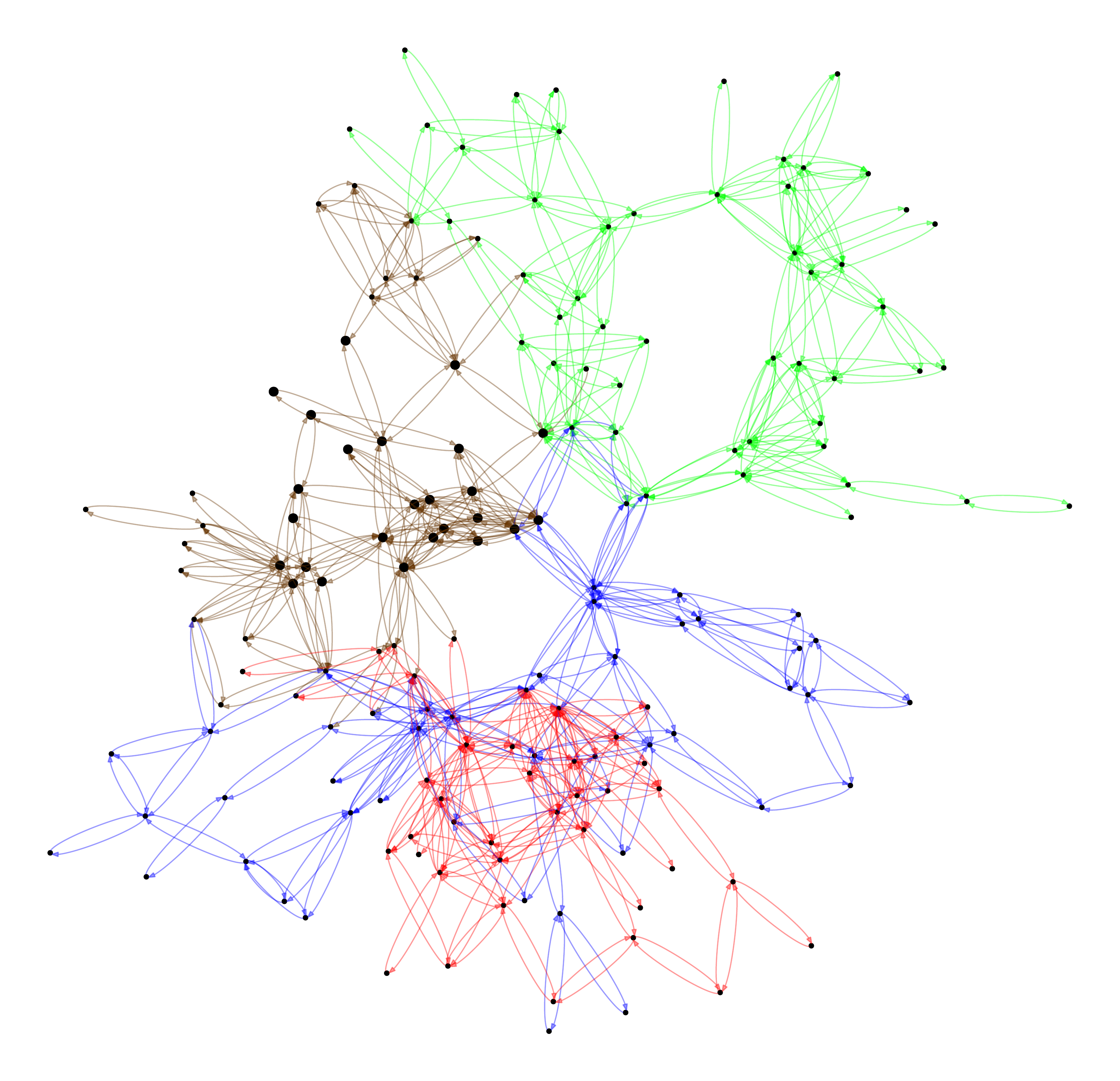}
\llap{\raisebox{0cm}{
      \includegraphics[angle=90,origin=c,trim = 14mm 8mm 45mm 8mm, clip,width=0.40\textwidth]{figures/shape_.png}%
    }}%
}\vspace{-0.5\baselineskip}
\caption{AS 1755 topology partitioned into four using uniform traffic distribution (left) and non-uniform traffic distribution (right). Details in text.}\label{fig:as1755}
\endminipage\hfill
\end{figure*}

%\begin{figure}
%\includegraphics[trim = 0mm 0mm 5mm 5mm, clip, width=1\columnwidth]{figures/1221_header_overheads.pdf}
%\caption{The expected bit length for a Bloom-filter that does not produce any false positives in XBF, and LIPSIN. The boxplot whiskers show 5th and 95th percentiles. The blue line shows the mean length for the XBF-filter}\label{fig:headerOverhead}
%\end{figure}

%\subsection{Improve XBF Scalability}
\subsection{Scalability of XBF}
\label{sec:eval:scalability}

We used the Barab\'{a}si--Albert (BA) and Erd\H{o}s--R\'{e}nyi (ER) models to evaluate how XBF scales as the network grows. The parameters we used with these models are summarized in Table~\ref{tbl:parametersSummary}.

Figure~\ref{fig:scalabilityPoppers} shows the number of popper switches through which a packet traverses as a function of network size. The growth in is clearly sub-linear. The sub-linearity comes largely from the fact that the network diameter grows sub-linearly -- the number of partitions on a path is limited by the network diameter. This guarantees the scalability of XBF in a very natural way.

The same sub-linear shape can also be seen in the number of popping operations (Figure \ref{fig:scalabilityPoppings}) and in the header overhead (Figure~\ref{fig:scalabilityHeaderOverhead}). However, the number of popping-operations (which approximates the processing overhead of the forwarding) grows slightly faster than the number of popper-switches. This is caused by the fact that as the network grows, the probability of a popper-switch having adjacent links belonging to more than two partitions grows. Nevertheless, both the growth in the popping operations and the growth of the header overhead are still sub-linear implying good scalability.

%A notable thing in figures~\ref{fig:scalabilityPoppers} and \ref{fig:scalabilityPoppings} is the growth in networks created with ER-model, which appears to be near linear. This is caused by the fact that the number of edges in ER-graphs is proportional to $|V|^2$ while in BA-model the number of edges grows linearly as a function of number of vertices. 

\subsection{Effect of the traffic patterns}
\label{sec:eval:trafficPatterns}

So far in the evaluation that every node sends equal amount of traffic to every other node. In order to gain understanding on how non-uniform traffic volumes may change the partitioning quality, we also ran the simulation with non-uniform traffic volumes. This simulation was executed as described in Section~\ref{sec:setup} with the difference that 10\% of the nodes were assigned as high-demand nodes, which produced ten times of the traffic that a normal node does.
%with the difference that the probability that a node with a random and unique identifier ${n \in^R \{1,\ldots,|V|\}}$ was chosen as a source or sink was proportional to ${P(n) \propto n^{-2}}$. 

The effect of this non-uniform traffic in AS 1221 topology can be seen in Figure~\ref{fig:poppingsSkewed}, which shows the mean number of popping-operations when the partitioning was done with Jigsaw-algorithm with and without taking the traffic volumes into account (the shaded areas indicate the standard deviation). The red stem-lines indicate the improvement when the traffic volumes were taken into account. As we suspected, the partitioning quality can be improved relatively much when the traffic patterns are non-uniform -- a typical improvement in this experiment was roughly 9\% in terms of the number of popping-operations when the traffic volumes are taken into account. 

It should be noted that we did not incorporate any kind of spatial preference to our simulation (i.e.\ a node would be more likely to communicate with nearby nodes). To illustrate what happens when nodes communicate mostly with nearby nodes, we repeated the experiment in AS 1755 topology so that the high-demand nodes were placed near to each other. Figure~\ref{fig:as1755} shows how the partitioning changes in this situation: when the traffic volumes are taken into account the partitioning algorithm places the high-demand nodes (under the shaded area) to the same partition. 

Based on these examples, we believe it is safe to say that results obtained assuming uniform traffic volumes show only the worst case performance of XBF and Jigsaw.

\subsection{Summary of the results}
\label{sec:eval:summary}

To summarise, XBF has been shown to be able to effectively reduce both header overhead and processing overhead in our evaluation. For conventional BF-based routing, both overhead are a function of total number of links which can grow fast in large networks. Whereas for XBF, by partitioning a network, such overhead are bounded by the network diameter (especially for small multicast trees) which grows much slower due to small-world effect. The reason is because a single packet will only traverse a small amount of partitions in the network. Although as the multicast tree grows, XBF eventually includes all the partitions in zBF, it still has significantly smaller overhead comparing to conventional BF-based routing.
%%%
Moreover, XBF is able to exploit traffic pattern when applying Jigsaw to further optimise its performance. Traffic awareness guarantees the volume of cross-partition traffic is minimized.

\section{SDN Integration}
\label{sec:sdn}

%%% So far we have not described the XBF-header in very detailed way. This has been intentional as our main focus is on the overall forwarding scheme rather than on  the exact composure of XBF header. It, however, would be ignorant to overlook the importance of this kind of details. Hence, we now describe the details of the protocol header as well as our prototype implementation.
%%%\lw{present in a way: complexity, work flow, key data structures, key optimisations, pros \& cons ...}

So far, we have been focusing on describing the overall XBF architecture and understanding its system behaviours. However, for a practical deployment, it is equally important to understand the complexity and flexibility in the realistic system implementation. %Hence, we now present the details of our prototype implementation.
%
%We 
Thus, we implemented the XBF popper and forwarder in P4 and verified the correctness of the scheme with emulations using the \emph{bmv2} P4 software switch~\cite{p4bmv2}. We chose P4 as the implementation language since it is currently one of the most promising SDN technology allowing packet processing on arbitrary fields. 
Our P4 implementation also serves as a sanity check that the packet processing does not include any operations that would be difficult to implement with hardware. The P4 implementation is about 200 lines of code indicating its low complexity.

%\red{This could be moved to the section where we describe the XBF protocol.}
%Figure~\ref{?} shows the format of a XBF header. 
In our implementation, the XBF header is within a IPv6 header, which contains the iBF in the source and destination address fields. The XBF header itself contains two main parts: a \emph{filter bitmap} and a \emph{partition Bloom-filter set}. The filter bitmap has a length of $|P|$ (number of partitions in the network) and tells which partition filters are included in the partition Bloom-filter set zBF (the bitmap has a 1-bit in position $n$ iff the multicast tree passes the partition $n$). The partition Bloom-filters are included in the XBF header in ascending order based on the partition identifier selected from $\{1,\ldots,|P|\}$. This makes it is possible to calculate the offset of any partition Bloom-filter using the information found from the filter bitmap. %Furthermore, this design makes it possible to compress sparsely filled zBFs. For example, it is possible to double the length of the filter-bitmap so that every bit refers to a 128-bit part of the zBF. This can be done independently of the partition size and may, in some situations, decreases the header overhead.

\begin{figure}[t]
	\includegraphics[trim = 5mm 5mm 5mm 5mm, clip, width=1\columnwidth]{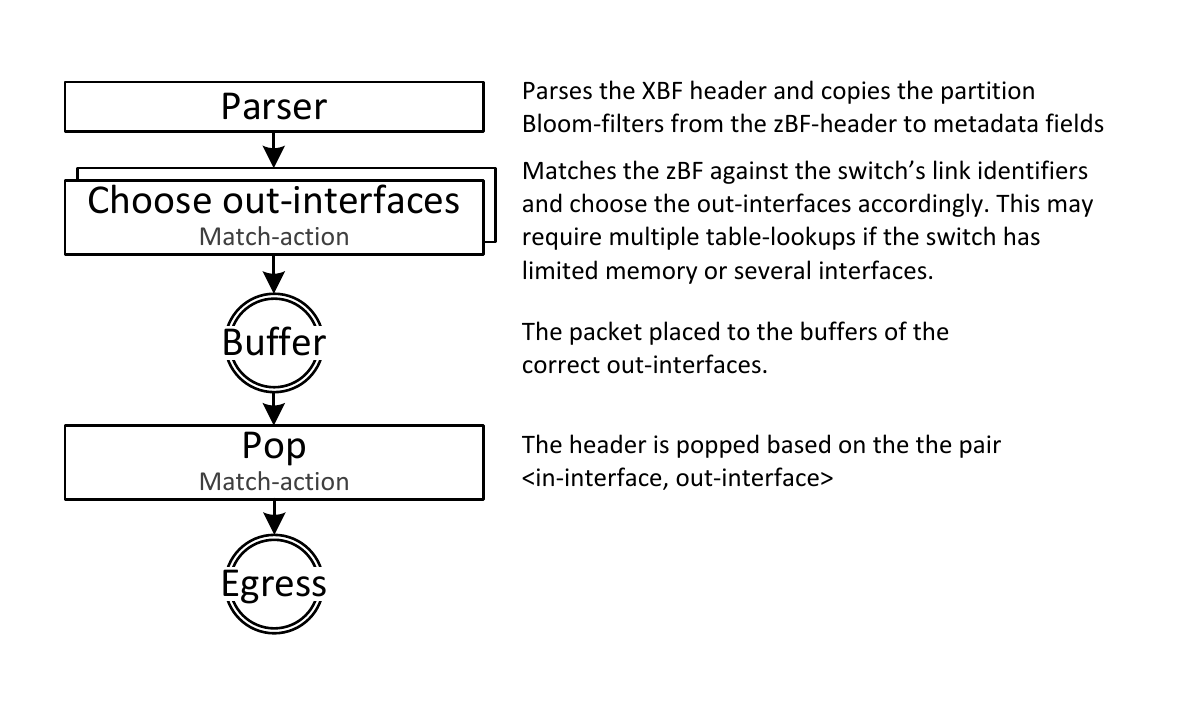}
	\caption{P4 implementation of the popper and the relevant parts of the switch's processing pipeline.}
	\label{fig:p4implementation}
\end{figure}

The pipeline of the P4 popper-switch is illustrated in Figure~\ref{fig:p4implementation}. The pipeline starts with parsing the XBF packet and extracting the partition Bloom-filters from the zBF. The parsing is a simple operation done with a state machine. When the packet has been parsed, the out-interfaces are chosen based on the values in the zBF headers (i.e., the popper does not have to check the Bloom-filter in the IPv6 header at all). Choosing the correct out-interfaces may require several table-lookups if the switch has several interfaces and does not support large tables -- a switch with 8 interfaces can choose the correct out-interfaces with one lookup on a table with 256 entries. 
%Since Internet traffic has been shown to possess strong temporal and spatial locality \cite{?} \ma{Liang: do you know a reference that fits here?}, it is highly likely the subsequent packets are from the same multicast tree, hence a small LRU cache can be adopted to further reduce processing overhead by avoiding unnecessary decompression. \lw{what is the key in the cache? zBF is too big to be a key ... if we use hash ... what is the complexity comparing to decompression algorithm ...}
%%%The LRU cache uses zBF as its key and checks if 

When the out-interfaces have been chosen and written to the packet metadata, the packet is replicated to the corresponding egress-interface queues. In the egress part of the pipeline, the switch performs a lookup based on the pair $\langle \textit{in-interface}, \textit{out-interface} \rangle$. As a result of this lookup, the packet is either popped or forwarded without modification to the egress queue.

%As discussed, popping is performance-wise the main cost in XBF. In our P4 implementation, the popping is done by sending the packet back to the beginning of the processing pipeline for every popping operation that is to be performed. The rest of the pipeline constituted of copying the partition Bloom-filter of the particular partition to the iBF and cloning the packet to the correct egress queues.

%While the popper implementation details are highly dependent on the target hardware, the protocol itself turned out to be extremely simple.

We acknowledge, that the details for an efficient XBF implementation are highly depended on the target hardware. However, our P4 prototype shows that XBF can be implemented with relative ease to very simple devices. The only modification we had to implement on the P4 virtual-switch \emph{bmv2}, which we used for testing, was to change the way how multicast groups are interpreted. 
%\lw{any chance to have some numbers for potential throughput?}

\section{Discussion and future work}\label{sec:discussion}

%\textbf{Consequences of the zero false-positive forwarding:}

%\ma{Can we figure out anything more to say about this?}

\textbf{On the false-positive-free forwarding:}
The design of XBF, as described in this paper, requires that the fixed-length partition Bloom-filter have at least as many bits as there are links in the partition. Because of this design choice, XBF does not suffer from false positives. It, however, should be noted that XBF-like partitioning approach works also with conventional Bloom-filter based forwarding system that allows false positives. However, we believe that the false-positive-free approach described in this paper is more palatable for two reasons.
%Even though XBF-like partitioning approach would work also with conventional Bloom-filter based forwarding system that allows false positives, we believe that the false-positive-free approach described in this paper is more palatable for two reasons:

%First of all, allowing false positives in multicast forwarding is questionable. The whole point of using multicast transmission over several unicast transmissions is to reduce the overall bandwidth consumption. Thus, we believe that the false positives should be avoided if possible. 

First of all, probabilistically occurring false positives make network management and debugging very difficult. 
%allowing false positives in multicast forwarding is questionable. The whole point of using multicast transmission over several unicast transmissions is to reduce the overall bandwidth consumption. Thus, we believe that the false positives should be avoided if possible. 
Second, by removing the possibility for false-pos\-itives, XBF solves the problems related to \emph{forwarding anomalies} that have plagued Bloom-filter forwarding~\cite{5935060}. As an example, it is possible that a false positive takes the packet to an earlier point in the multicast tree. This kind of situation would effectively cause an infinite loop in the network. False-positive-free XBF does not suffer from these problems. There are several existing solutions to solve the problems caused by forwarding anomalies which, however, all increase the forwarding complexity. (We refer the reader to S\"arel\"a et al.\ for a more comprehensive description about forwarding anomalies~\cite{5935060}). 

\textbf{On the addressing:}
XBF is basically an addressing scheme where every link in the network is identified with a unique bit in a bitsring. It may appear couterintuitive that embedding this information into every packet is feasible. This, nevertheless, is the case as we have shown. The key insight is that the links in a multicast tree are not independent from each other. Thus, the network partitioning becomes an effective way to compress the information embedded into the packets.

In a more abstract sense, XBF can also be described as a traffic aware addressing scheme -- if there is a lot of traffic between two nodes, they are more likely to be put in a single partition. We are not aware of any previous work that would have suggested optimising the addressing based on the network traffic patters.

\textbf{Adapting to changing traffic patterns:}
In Section~\ref{sec:forwarding}, we claim that the topology manager would perform the partitioning only once. However, traffic patterns may change over time thus rendering the partitioning ineffective. In fact, diurnal mobility patterns and changes in the traffic volumes are often seen in real networks~\cite{fowler1991local}. Because of this, there should be some way how the partitioning can be changed without disrupting the ongoing communication -- naive repartitioning would render the previously used Bloom-filters invalid causing extra delays.

The solution to this is to allow several coexisting partitionings. The idea is to use some header field, such as IPv6-header's traffic-class field, to identify a ``partitioning'' that a multicast tree uses. Obviously two multicast trees that use different partitioning cannot be combined. Nevertheless this mechanism could be used to reduce the overhead of the repartitioning. In addition to the signalling cost, the only drawback of this is that the flow-tables would grow $n$-fold where $n$ is the number of coexisting partitionings.

\textbf{Handling topology changes:}
Although Bloom-filter forwarding is intended for relatively static networks, no network is immune to occasional topology changes. In general, XBF handles changes in the network topology as any other Bloom-filter forwarding scheme. That is, if one or more links disappear from the network (e.g.\ due to a switch malfunction), these link-identifiers are no longer used when constructing paths. Moreover, when new links are introduced to the network, the topology manager simply assigns them new link-identifiers (this may require creating an entirely new partition). If this leads to non-optimal partitioning, the network can be repartitioned, as described above, once the topology has remained static for certain time.

Due to the fact that the topology changes can be handled as done in other Bloom-filter forwarding schemes, we refer the reader to~\cite{zahemszky2009fast} for more information about this matter.

%\as{also what is the overhead for the topology manager for change in network topology? can you cite some existing SDN related work for this? and say that this is also applicable for xbf - else the reviewer may end up asking this question asking us why we didn't consider the topology changes and the overhead it causes to the centralised topology manager..}

%The only real drawback of the proposed false-positive-free forwarding scheme is that it slightly increases the header overheads of unicast transmissions in networks that have a high diameter. Although we show in Section~\ref{sec:results} that the header overhead is reasonable, it still may be too large for certain scenarios. However, there are several other ways how the headers can be made smaller. First, the zBF could be compressed. Compressing just the zBF would be palatable as it only affects the performance of the poppers. The zBF would also be extremely compressible due to the fact that it is mostly zeros. Another way how the header sizes can be made smaller is to reuse some of the link-identifiers. For example, it is possible to use the one link-identifier for a bidirectional link without causing any false forwarding positives if the network does not utilize multi-path forwarding. This simple trick would effectively double the partition sizes and thus greatly decrease the header overhead.

%\textbf{Packet fragmentation and segmentation:}
%\ma{Something about the header becoming too large and packet fragmentation.}

\textbf{Security:}
Bloom-filter forwarding has several well known security problems. In general, it should not used in untrusted environments without augmenting the protocol with hop-by-hop security checks that would significantly increase the forwarding complexity~\cite{6616021}. XBF, however, does not have this problem as it is intended for intra-domain use, where all nodes are trusted. Extending the work for inter-domain forwarding is left as a future exercise.

%\subsection{Other stuff}
%\lw{Can we drop the corresponding BF in the header if the packet leaves the zone? Do we really have the re-entering issue in reality?}
%\ma{I dislike this idea. Re-entering probably happens very rarely when using shortest path routing. However, there are legitimate reasons for not using shortest paths (e.g. multipath routing or routing via a cache). Also, dropping BF-headers decreases the generality of the forwarding mechanism without bringing significant benefits.}
\section{Conclusion}

%We conclude that XBF is \emph{awesome} and should be deployed in every network in the world.

In-packet Bloom-filter forwarding was originally advocated as simple and fast way to do multicast forwarding. While this promise holds with small networks and multicast groups, the technology does not scale well -- if the network or multicast groups become too large, the forwarding efficiency decreases due to false positives. The previously proposed fixes to this scalability problem have mainly relied on varying the length of the in-packet Bloom-filter. These solutions, however, increase the forwarding complexity significantly effectively countering one of the original design goals which was to enable simple and fast multicast forwarding. 

The XBF-forwarding scheme proposed in this paper fills this gap. It makes Bloom-filter forwarding scalable by partitioning the network into equisized partition that reflect the network topology and traffic patterns, and uses a separate fixed-length Bloom-filter in each of these. Because of this, the forwarding inside every partition is kept very simple regardless of the topology or multicast-group size. Also, the partitioning also allows us to to eliminate the false-positives from Bloom-filter-based forwarding. Our evaluation shows that XBF is implementable with the existing SDN technologies and scales to very large networks.

\begin{comment}
\section*{\red{Markku's notes}}
\textbf{Add following thought to some place}

The issue with many previous Bloom-filter based research is that they have used false-positive ration as the metric to be optimized. The problem with fpr is that it only gives the probability that a false positive occurs during a single membership test. Thus, different topologies (or multicast trees) with the same fpr may produce different number of false positives because there number of membership tests may vary between these.

The fact that fpr does not yield any information about the underlying topology is problematic.
\end{comment}

\bibliographystyle{IEEEtran}
\bibliography{markku}
\end{document}